\documentclass[iop, apj]{emulateapj}
\usepackage{threeparttable}
\usepackage{multirow}
\usepackage{times}
\usepackage{enumitem}
\usepackage{url} 
\usepackage{color}
\usepackage{amsmath,bm}
\usepackage{float}

\shorttitle{Photometric Metallicities and Distances for $\sim$ 136,000 RR Lyrae Stars}
\shortauthors{ Xin-Yi Li et al.}

\begin{document}

\title{photometric metallicity and distance estimates for $\sim$ 136,000 RR Lyrae stars from Gaia DR3}

\author{Xin-Yi Li\altaffilmark{1}}
\author{Yang Huang\altaffilmark{2,3,1,8}}
\author{Gao-Chao Liu\altaffilmark{4}}
\author{Timothy C. Beers\altaffilmark{5}}
\author{Hua-Wei Zhang\altaffilmark{6,7}}

\altaffiltext{1}{South-Western Institute for Astronomy Research, Yunnan University, Kunming 650500, People's Republic of China;}
\altaffiltext{2}{School of Astronomy and Space Science, University of Chinese Academy of Sciences, Beijing 100049,  People's Republic of China, {\it huangyang@bao.ac.cn {\rm (YH)}};}
\altaffiltext{3}{Key Laboratory of Space Astronomy and Technology, National Astronomical Observatories, Chinese Academy of Sciences, Beijing 100101, People's Republic of China}
\altaffiltext{4}{Center for Astronomy and Space Sciences, China Three Gorges University, Yichang 443002, People's Republic of China}
\altaffiltext{5}{Department of Physics and Astronomy and JINA Center for the Evolution of the Elements, University of Notre Dame, Notre Dame, IN 46556, USA}
\altaffiltext{6}{Department of Astronomy, Peking University, Beijing 100871, People's Republic of China}
\altaffiltext{7}{Kavli Institute for Astronomy and Astrophysics, Peking University, Bejing 100871, People's Republic of China}
\altaffiltext{8}{Corresponding author}

\begin{abstract}
We present a sample of 135,873 RR Lyrae stars (RRLs) with precise photometric-metallicity and distance estimates from our newly calibrated $P$--$\phi_{31}$--$R_{21}$--[Fe/H]/$P$--$R_{21}$--[Fe/H] and $G$-band\,absolute\,magnitude-metallicity relations. The $P$--$\phi_{31}$--$R_{21}$--[Fe/H] and $P$--$R_{21}$--[Fe/H] relations for type RRab and type RRc stars are obtained from nearly 2700 {\it\,Gaia}-identified RRLs with precise $\phi_{31}$ and $R_{21}$ measurements from the light curves and metallicity estimates from spectroscopy. Using 236 nearby RRLs with accurate distances estimated from parallax measurements with {\it\,Gaia} EDR3, new $G$-band\,absolute\,magnitude-metallicity relations and near-infrared period-absolute magnitude-metallicity relations are constructed. External checks, using other high-resolution\,spectroscopic sample of field RRLs and RRL members of globular clusters, show that the typical uncertainties in our photometric-metallicity estimates are about 0.24\,and\,0.16\,dex for type RRab and type RRc stars, respectively, without significant systematic bias with respect to high-resolution\,spectroscopic metallicity measurements. The accuracies of these metallicity estimates are much improved, especially for type RRab stars, when compared to those provided by the {\it\,Gaia} DR3 release. Validations of our distance estimates, again by using members of globular clusters, show that the typical distance errors are only 3--4\%. The distance modulus $\mu_0=18.503\pm0.001(stat)\pm0.040(syst)$\,mag for the Large Magellanic Cloud (LMC) and $\mu_0=19.030\pm0.003(stat)\pm 0.043(syst)$\,mag for the Small Magellanic Cloud (SMC) are estimated from our RRab star sample, respectively, and are in excellent agreement with previous measurements. The mean metallicities of the LMC and SMC derived in this work are also consistent with the previous determinations. Using our sample, a steep metallicity gradient of $-0.024\pm0.001$\,dex\,kpc$^{-1}$ is found for the LMC, while a negligible metallicity gradient is obtained for the SMC.

\end{abstract}
\keywords{RR Lyrae variable stars; Metallicity; Distance; Globular clusters; Magellanic clouds}

\section{Introduction}
RR Lyrae stars (RRLs) are old ($>$10\,Gyr), low-mass ($<1$\,M$_\odot$) metal-poor periodic pulsating variable stars, mostly distributed in the bulge, thick disk, globular clusters, stellar halo, and substructures (e.g., dwarf galaxies and stellar streams) within the Galaxy. 
They are core helium-burning stars located on the horizontal branch on the Hertzsprung-Russell diagram, with relatively bright luminosities ($M_V \sim$\,0.65\,mag; e.g., \citealt{2008ApJ...676L.135C,2018MNRAS.481.1195M}). According to their pulsating modes, RRLs can be divided into three types: RRab -- fundamental-mode pulsating stars; RRc -- first-overtone pulsating stars; and RRd -- double-mode pulsating stars. They are ideal standard candles, thanks to their well-defined absolute magnitude-metallicity relations in visual bands and period-absolute magnitude-metallicity relations in the near/mid-infrared bands. The above advantages of RRLs make them excellent tracers to probe the properties of our Galaxy, especially the stellar halo \citep{2013ApJ...763...32D,2019MNRAS.482.3868I,2021MNRAS.502.5686I,2022MNRAS.513.1958W}. 

Over the years, many large-scale time-domain surveys have been conducted, and  released large samples of RRLs, such as the Pan-STARRS1 survey (\citealt{2017AJ....153..204S}), the Optical Gravitational Lensing Experiment (OGLE, \citealt{2019AcA....69..321S}), the Catalina survey (\citealt{2013ApJ...765..154D, 2014ApJS..213....9D}), the All-Sky Automated Survey for SuperNovae (ASAS-SN, \citealt{2019MNRAS.486.1907J}), and the $Gaia$ Mission (\citealt{2018A&A...618A..30H,2019A&A...622A..60C,2022arXiv220606278C}). 

Compared to photometric surveys, spectroscopic observations of RRLs are quite limited. Such spectra are particularly valuable, as they can provide vital measurements of RRLs, such as atmospheric parameters (effective temperature, $T_{\rm eff}$, surface gravity, log\,$g$, and metallicity, [Fe/H]), and line-of-sight velocity ($v_{\rm los}$), to enable studies, not only of the structure of the Milky Way, but also of the chemical and kinematic properties of our Galaxy. However, the measurements of stellar parameters for RRLs are quite challenging, since they are pulsating on short time-scales (0.2-1.0 days), and their spectra can vary with time even during a single exposure. 

The most accurate way to determine the stellar parameters of RRLs is using high-resolution spectra observed at suitable phases (to avoid the effects from shock waves; e.g., \citealt{1992MNRAS.256...26F,2015MNRAS.447.2404P}). But only a few hundred RRLs have been observed in this way (e.g., \citealt{2011ApJS..197...29F, 2012MNRAS.422.2116K, 2013ApJ...773..181N, 2014ApJ...782...59G, 2015MNRAS.447.2404P}), due to the limited time available on large-aperture telescopes. 

To obtain reliable metallicity estimates for a large sample of RRLs, Preston (1959) first proposed the $\Delta S$ index method for low-resolution spectra observed at the phase of minimum light. Taking advantage of massive low-resolution spectroscopic surveys, \citet[][ hereafter L20]{2020ApJS..247...68L} present the largest catalog of over 5000 RRLs with metallicity and $v_{\rm los}$ measurements obtained with a template-matching method. Metallicities of RRLs can be accurately estimated down to about [Fe/H] = $-3.0$ by this method, with a typical uncertainty of 0.2\,dex when compared to the results of high-resolution spectroscopy. This large catalog of RRLs is very powerful for studying the global properties and substructures of the Galactic stellar halo \citep{2022MNRAS.517.2787L,2022MNRAS.513.1958W}. 

Compared to the RRL samples from photometric surveys, RRLs with metallicity information built upon large-scale spectroscopic surveys also have significant shortcomings -- e.g., shallower limiting magnitudes (thus smaller volumes covered)  and sparse sampling (thus suffering from potential selection effects). 
Given the pulsation nature of RRLs, their light curves are decided by few physical parameters, including their chemical compositions.
Observationally, such relations between metallicity and the parameters yielded by decompositions of the light curves of RRab stars are found by \cite{1992ApJ...395..192C,1993ApJ...412..183C} and \cite{1995A&A...293L..57K}. Later, such relations have been applied to derive photometric metallicities for RRLs with precise light curve measurements (e.g., \citealt{1996A&A...312..111J,2005AcA....55...59S,2007MNRAS.374.1421M,2013ApJ...773..181N,2016ApJS..227...30N,2018ApJ...857...55H,2021ApJ...920...33D,2021MNRAS.502.5686I,2021ApJ...912..144M,2022ApJ...931..131M,2022ApJS..261...33D}). 

By using a few hundred local RRLs with very accurate light curves, and metallicity estimated from high-resolution spectroscopy, \cite{2013ApJ...773..181N} presented very tight relations between [Fe/H] and parameters derived  from light curves (i.e., the period and Fourier decomposition parameter $\phi_{31}$) for type RRab and type RRc stars, respectively, with scatters of about 0.1\,dex. Most recently, \citet[][hereafter C22]{2022arXiv220606278C} released a large sample of 270,905 RRLs with full sky coverage identified from the light curves obtained by $Gaia$ DR3. By transforming the relations built by \cite{2013ApJ...773..181N} from the Kepler $Kp$-band to the $Gaia$ $G$-band, C22 present metallicity estimates for 133,559 RRLs with precise period and $\phi_{31}$ measurements from $Gaia$ DR3. However, compared to the spectroscopic estimates, the metallicities derived by C22 have significant systematic bias and large uncertainties (especially for type RRab stars, see Section\,3 for a detailed discussion). 

In this study, making use of $\sim$2700 RRLs with precise metallicity estimates from L20, and  period, $\phi_{31}$. and $R_{21}$ measurements from the $Gaia$ $G$-band (C22), we calibrate the $P$--$\phi_{31}$--$R_{21}$--[Fe/H] and $P$--$R_{21}$--[Fe/H] relations\footnote{We note that C22 adopted the most often used $P$--$\phi_{31}$--[Fe/H] relations for both type RRab and type RRc stars. However, the comprehensive analysis by \citet{2021ApJ...920...33D} show that the $A_2$ term from Fourier decomposition is also tightly correlated with [Fe/H], especially for type RRc stars, and is even more important than the contribution from $P$ (see Figs. 2 and 3 of \citealt{2021ApJ...920...33D}).
We thus adopt a $P$--$\phi_{31}$--[Fe/H] relation for type RRab stars and a $P$--$R_{21}$--[Fe/H] relation for type RRc stars (no significant improvement is obtained by adding the $\phi_{31}$ term).
Here we use $R_{21}$ to represent $A_2$ since no direct measurements of the latter is provided in  the Gaia DR3 RRL catalog.} for type RRab and type RRc stars in the $Gaia$ $G$-band directly. 
With the metallicity estimated from the new relations for $\sim$ one thousand local RRLs with accurate distances estimated from parallax measurements, we further refine the $G$-band absolute magnitude-metallicity relations and the near-infrared period-absolute magnitude-metallicity ($PMZ$) relations in the $K_{\rm S}$- and  $W1$- bands for type RRab and type RRc stars, respectively. 


This paper is organized as follows. In Section\,2, we briefly describe the adopted data. In Section\,3, we present the $P$--$\phi_{31}$--$R_{21}$--[Fe/H] and $P$--$R_{21}$--[Fe/H] relations and the $M_{G}$--[Fe/H], $PM_{K_{\rm S}}Z$, and $PM_{W1}Z$ relations for RRLs, and derive metallicities and distances for the full set of RRLs with precise $\phi_{31}$ (and/or $R_{21}$) measurements based on the newly constructed relations. We present various checks on the derived photometric metallicities and distances in Section\,4. In Section\,5, we present the final RRL sample, and describe potential applications of this catalog. Finally, a summary is presented in Section\,6.

\section{Data}

\subsection{The $Gaia$ RRLs Sample}
Based on the data released in {\it Gaia} DR3 \citep{GaiaDR3}, C22 have published  270,905 RRLs (including 174,947 stars of type RRab, 93,952 stars of type RRc, and 2006 stars of type RRd) processed by the dedicated Specific Objects Study (SOS) Cep\&RRL pipeline \citep{2019A&A...625A..97R,2019A&A...622A..60C}.
This pipeline provides estimates of vital information for those RRLs derived from the light curves of {\it Gaia} DR3, including the pulsation parameters (period, epoch of maximum light, peak-to-peak amplitudes, and intensity-averaged mean magnitudes for the $Gaia$ $G$, $G_{\rm BP}$, and $G_{\rm RP}$ bands) and the Fourier decomposition parameters ($\phi_{21}$, $\phi_{31}$, and $R_{21}$) from the $G$-band light curves.
The SOS Cep\&RRL pipeline also delivers radial velocity information for 1100 bright RRLs. 
The pulsation period, $P$, and $G$-band amplitude are available for all 270,905 RR Lyrae candidates, and the $\phi_{31}$ and $R_{31}$ Fourier decomposition parameters are available for 135,873 RRLs. 
In addition, the $G$-band absorption $A_{G}$ of 142,660 type RRab stars are calculated from the $G$-band amplitude, pulsation period, and color, $G-G\rm_{RP}$ (see C22 for details). 
They also estimate the photometric metallicities for 133,559 RRLs with available values of $P$ and $\phi_{31}$.

\subsection{The RRL Sample with Spectroscopic Metallicity Estimates}
L20 presented a large sample of metallicity estimates for 5290 RRLs. Their metallicities are estimated by matching over 30,000 single-exposure low-to-medium resolution spectra collected from the LAMOST \citep{2012RAA....12..735D, 2012RAA....12..723Z, 2014IAUS..298..310L} and the SEGUE \citep{2009AJ....137.4377Y} surveys to the synthetic spectra.  Various tests show that the typical uncertainty of their estimated metallicity is about 0.2\,dex, without significant systematic offsets. 

To calibrate the $P$--$\phi_{31}$--$R_{21}$--[Fe/H] and $P$--$R_{21}$--[Fe/H] relations for type RRab and type RRc stars (see next section), the spectroscopic RRL sample of L20 is cross-matched with the photometric sample of C22, and 2687 RRLs in common (2046 type RRab and 641 type RRc stars; hereafter RRL\_CAL\_META sample) by appling the following cuts: 1) spectral signal-to-noise ratios greater than 20; 2) measurement errors of [Fe/H] smaller than 0.2/0.15\,dex for type RRab/RRc stars; 3) measurement errors of $\phi_{31}$ smaller than 0.5 for type RRab stars; 4) and measurement errors of $R_{21}$ smaller than 0.15/0.05 for type RRab/RRc stars. Figure\,1 presents the sky distribution of these stars.

\begin{figure*}
\begin{center}
\centering
\includegraphics[width=6.7in]{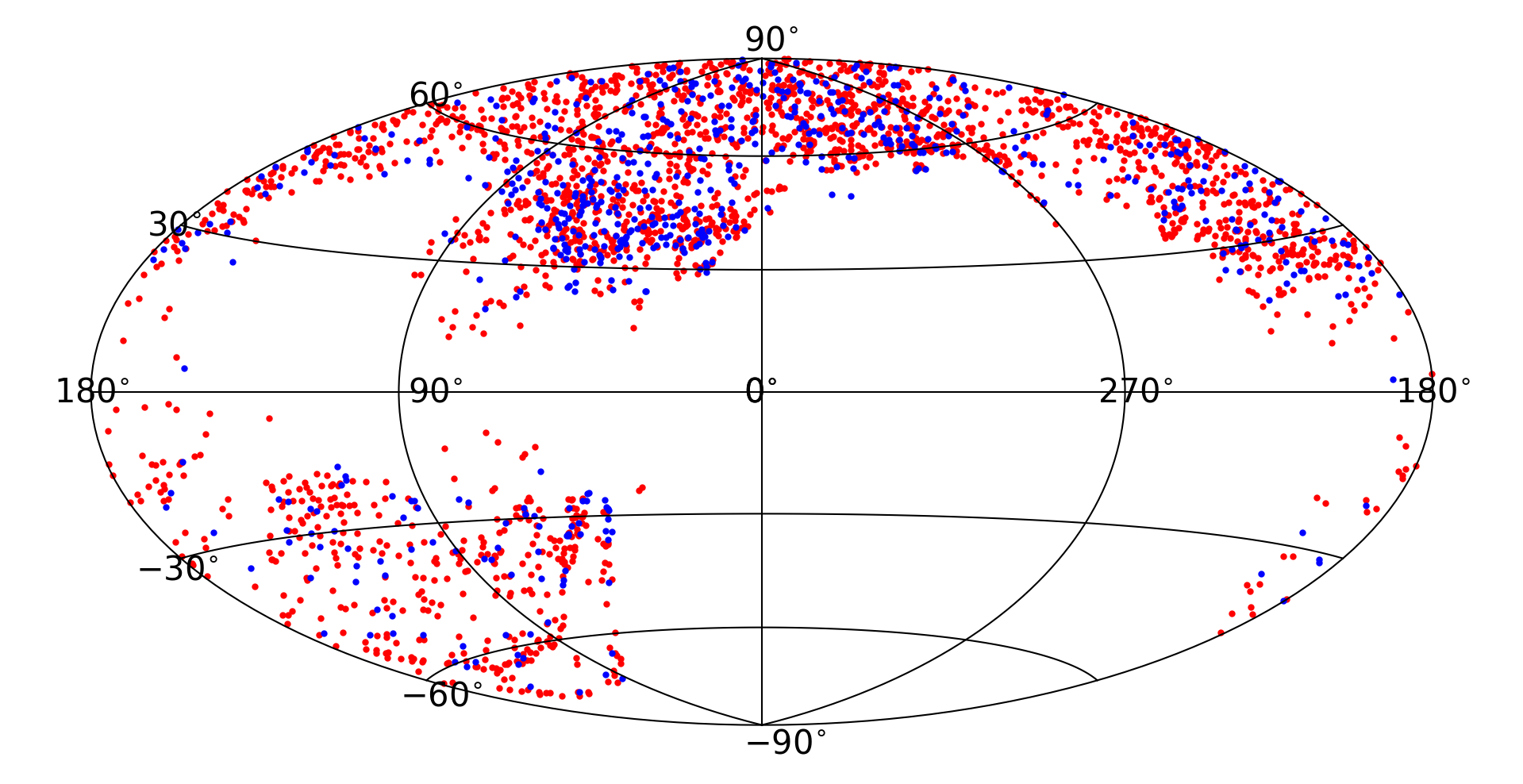}
\caption{Sky distribution of our RRL\_CAL\_META sample of 2046 type RRab (red dots) and 641 type RRc stars (blue dots) in Galactic coordinates.}
\end{center}
\end{figure*}

\begin{figure*}
\begin{center}
\includegraphics[width=6.7in]{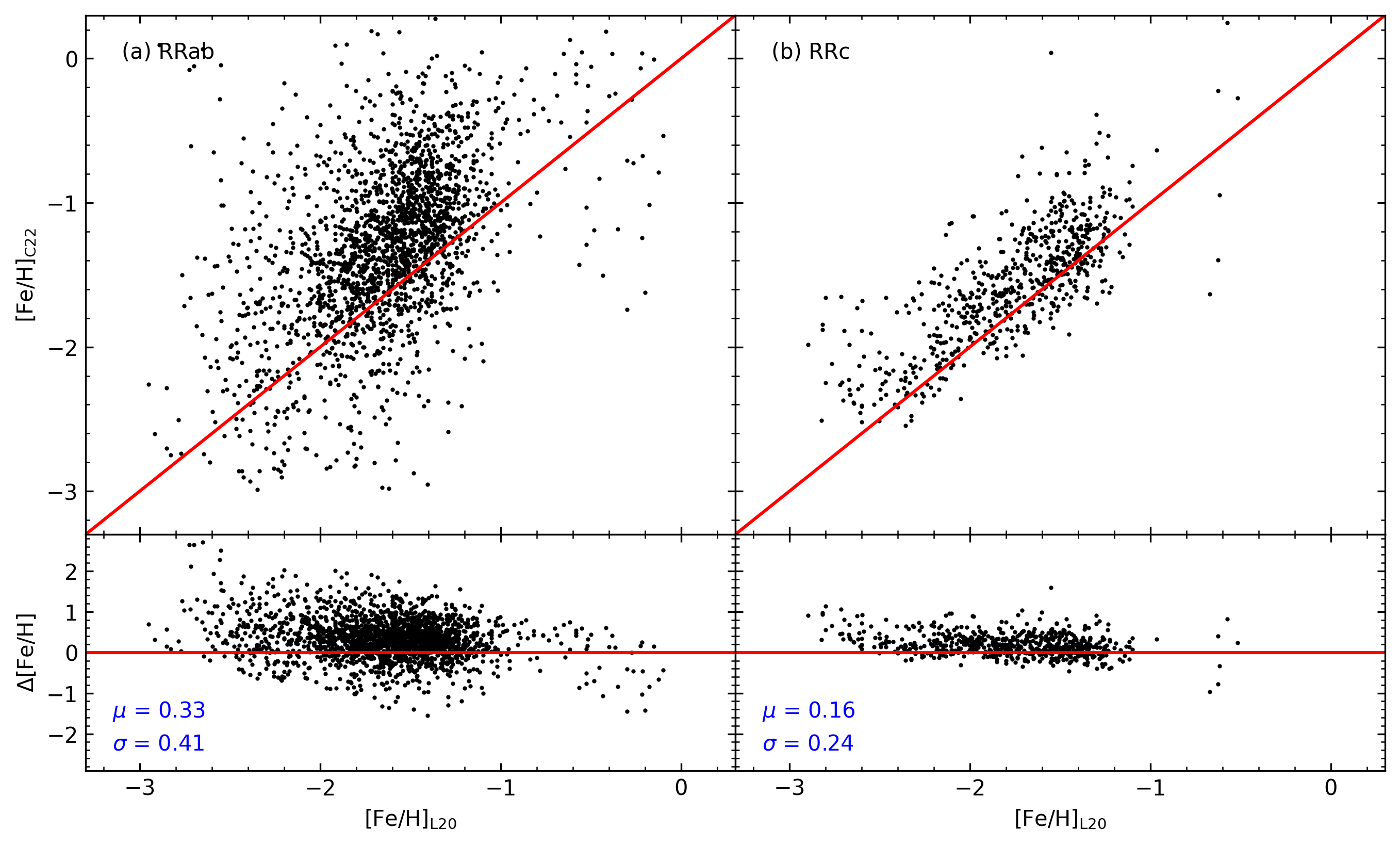}
\caption{Comparison of the estimated photometric metallicity from Clementini et al. (2022), [Fe/H]$\rm_{C22}$, with the spectroscopic metallicity, [Fe/H]$\rm_{L20}$, from Liu et al. (2020) for 2069 type RRab and 641 type RRc stars, respectively. The differences in metallicities, {$\rm\Delta[Fe/H]$} (in the sense C22 minus L20) are shown in the lower part of each panel, with the mean and standard deviation marked in the bottom-left corner.}
\end{center}
\end{figure*}

\begin{figure*}
\begin{center}
\includegraphics[width=6.7in]{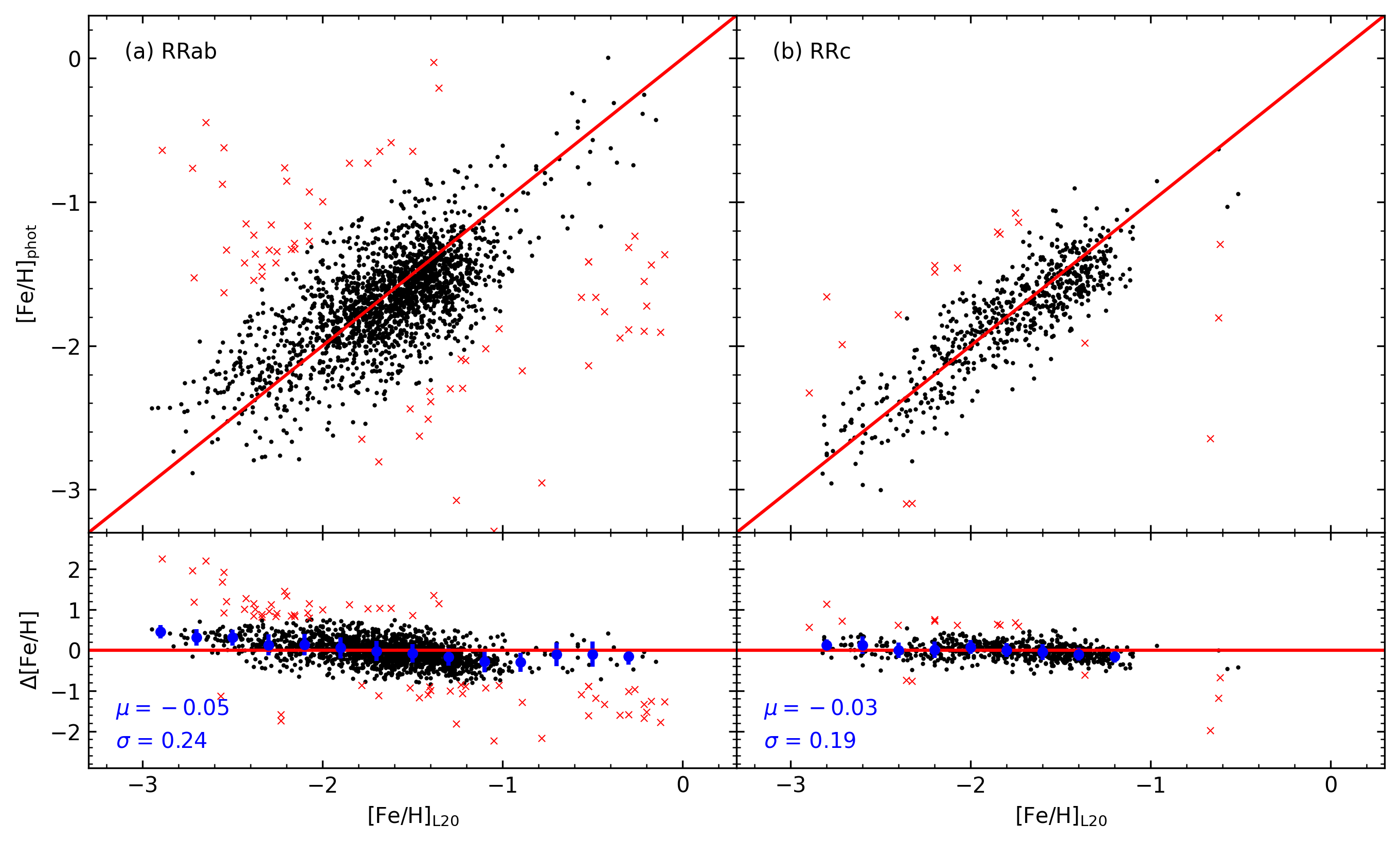}
\caption{{\it Left panel}: The top shows spectroscopic and our photometric metallicities for 2046 type RRab stars in our RRL\_CAL\_META sample. {\it Right panel}: Same as the left panel, but for 641 type RRc stars in our RRL\_CAL\_META smaple. In both panels, red crosses are data points excluded by 3$\sigma$ clipping, and black dots represent those adopted ones in the last fitting. The differences {$\rm\Delta[Fe/H]$} (in the sense this work minus that of L20) are shown in the lower part of each panel, with the mean and standard deviation marked in the bottom-left corner. The blue dots and associated error bars are the median values and standard deviations of {$\rm\Delta[Fe/H]$} in the individual [Fe/H]$_{\rm L20}$ bins.}
\end{center}
\end{figure*}

\begin{figure}[!htbp]
\begin{minipage}[t]{0.5\linewidth}
\centering
\includegraphics[width=3.2in]{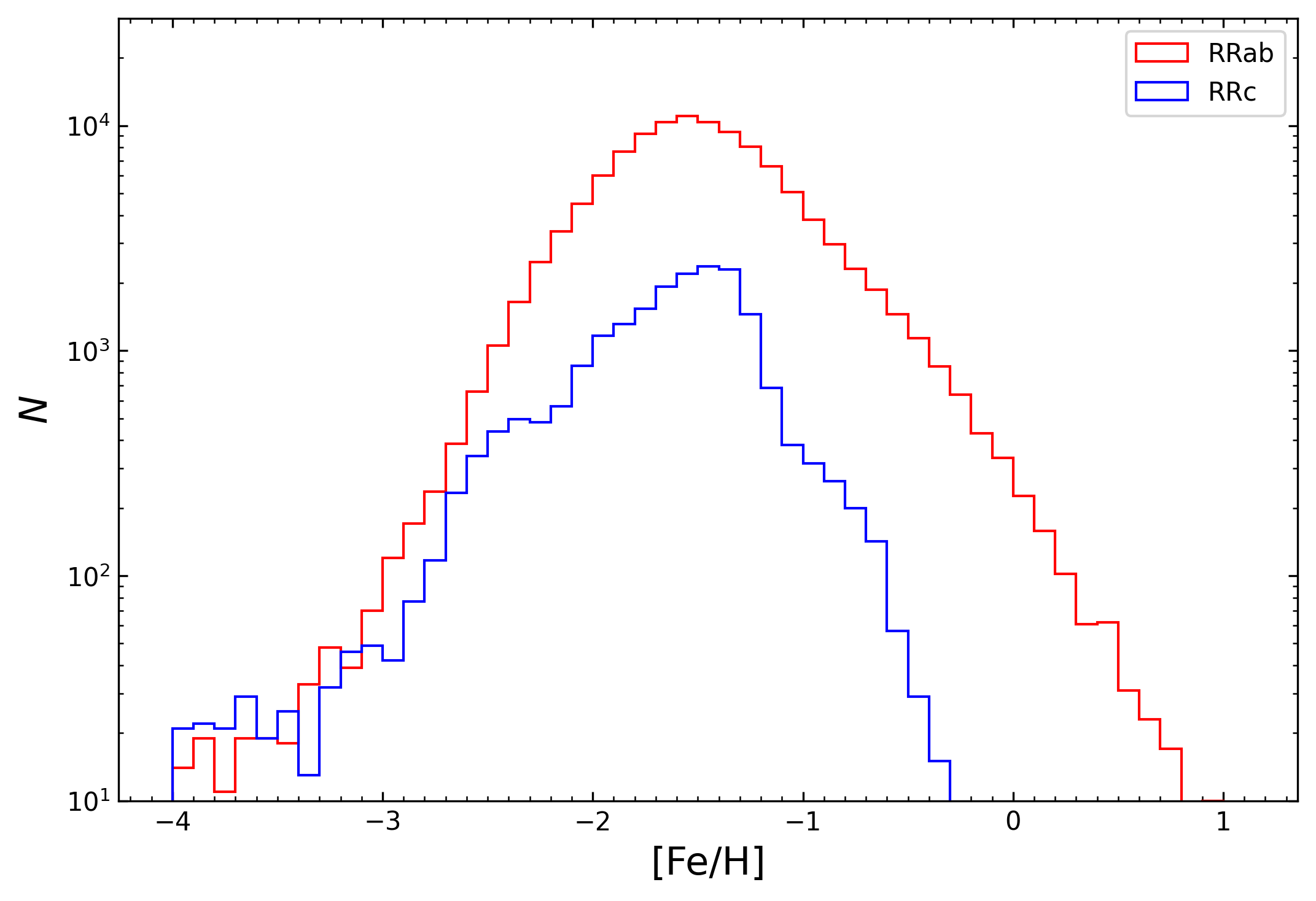}
\end{minipage}
\caption{Photometric-metallicity distributions of type RRab (red line) and type RRc stars (blue line) in our final RRL sample.}
\end{figure}

\begin{figure*}
\begin{center}
\centering
\includegraphics[width=6.7in]{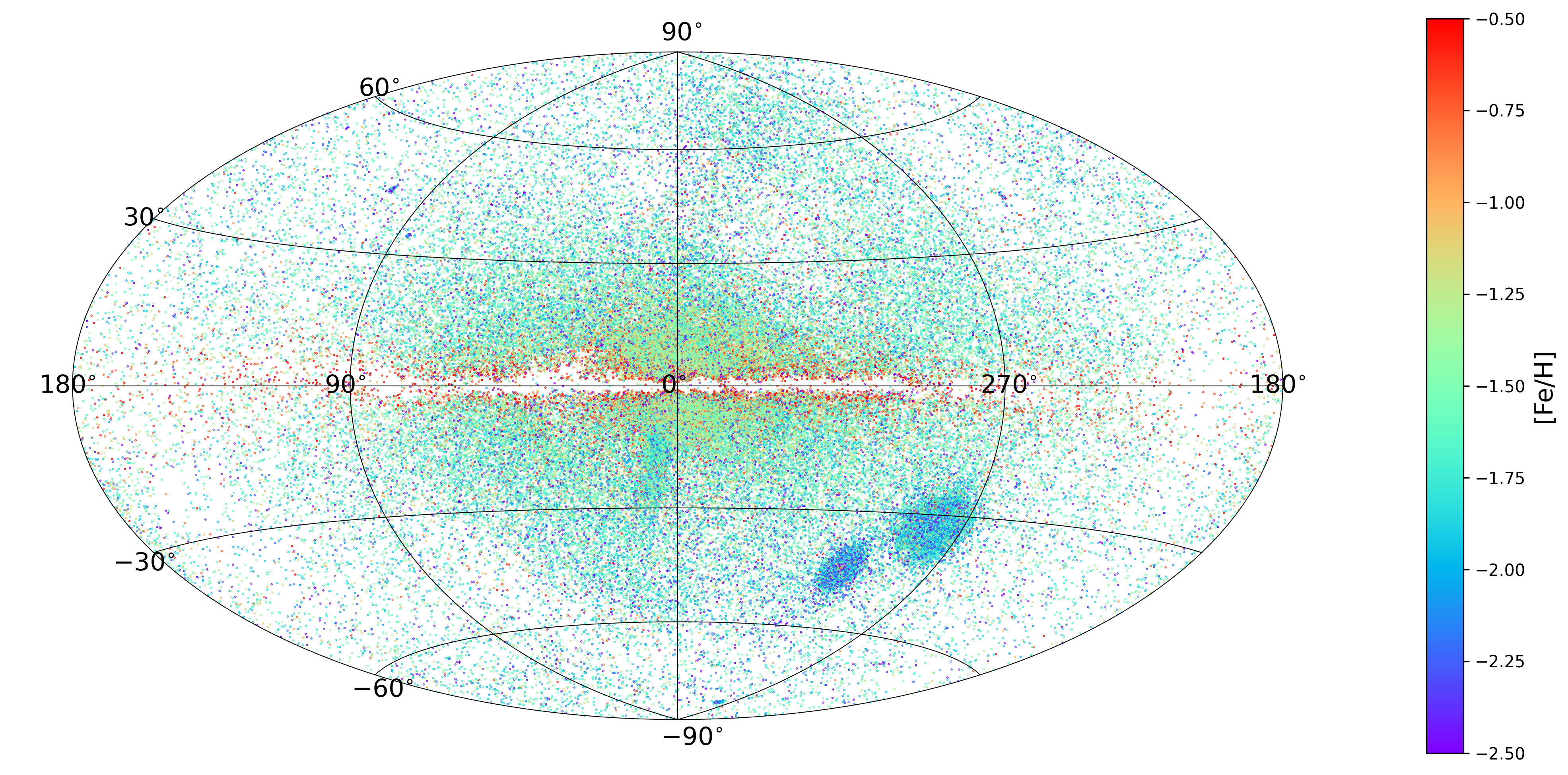}
\caption{Sky distribution of our final RRL sample of 135,873 RRLs (115,410 type RRab and 20,463 type RRc stars) in Galactic coordinates, color-coded by metallicity, as shown in the color bar.}
\end{center}
\end{figure*}

\section{Metallicity and Distance Estimates}
In this section, we calibrate the $P$--$\phi_{31}$--$R_{21}$--[Fe/H] and $P$--$R_{21}$--[Fe/H] relations for type RRab and type RRc stars, respectively, by using the RRL\_CAL\_META sample described in the previous section.  
These relations are then applied for all the RRLs with precise measurements of $P$, $\phi_{31}$, and $R_{21}$ to derive their metallicties. 
Moreover, using about one thousand local bright RRLs with accurate distances from $Gaia$ parallax measurements and photometric metallicities, the $M_{G}$--[Fe/H] and $PMZ$ relations for RRLs are constructed. The newly derived $M_{G}$--[Fe/H] relations are then used to derive distances of all RRLs with photometric-metallicity estimates.

\subsection{Metallicity}
Before calibrating the new $P$--$\phi_{31}$--$R_{21}$--[Fe/H] and $P$--$R_{21}$--[Fe/H] relations, we compare the photometric metallicity estimates to those from C22 using the RRL\_CAL\_META sample. The results are shown in Figure\,2. For type RRab stars, the photometric metallicities from C22 are systematically higher than those of the spectroscopic estimates of L20 by 0.33\,dex, along with a dispersion of 0.41\,dex. For type RRc stars, the photometric metallicties of C22 are slightly higher than those of L20, by 0.16\,dex, with a scatter of 0.24\,dex. 
 
To improve the accuracies of the photometric-metallicity estimates, we calibrate the $P$--$\phi_{31}$--$R_{21}$--[Fe/H] and $P$--$R_{21}$--[Fe/H] relations for type RRab and type RRc stars, respectively, using the RRL\_CAL\_META sample as described above.

For type RRab stars, the relation is built by fitting data points to the linear model inspired by the comprehensive analysis of \cite{2021ApJ...920...33D}:
\begin{equation}
\begin{split}
{\rm[Fe/H]} =\,& a_{0} + a_{1}(P-0.6) + a_{2}(\phi_{31}-2) \\  &+ a_{3}(R_{21}-0.45) \text{,}
\end{split}
\end{equation}
where $P$ is the period, $\phi_{31}$ and $R_{21}$ are the Fourier parameter of the $Gaia$ $G$-band light curve, [Fe/H] is the spectroscopic metallicity estimated by L20, and $a_{i}$ ($i$ = 0, ..., 3) are the fit coefficients. 

As shown in Figure\,3, the metallicities of our RRL\_CAL\_META sample are not uniformly 
distributed, with most of the stars located between [Fe/H] = $-1.0$ and $-2.0$, and fewer stars 
outside of this range, which can lead to biases in fitting the relation.
Similar to \cite{2022ApJS..261...33D}, we have applied density-dependent sample weights to the fitting process. 
The weight ($\omega_d$) of each star is calculated from the normalized Gaussian kernel density $\rho_d$ of the metallicity distribution.
We set $\omega_d$  to $0.6/\rho_d$ for $\rho_d \ge 0.15$, and to a constant value of 4 if $\rho_d < 0.15$.
The latter threshold is introduced to avoid the excessive influence from Possion noise due to fewer data points in deriving the density.
Three-sigma clipping is also performed in the fitting process. 

The resulting fit coefficients are listed in Table\,1.  A comparison between the spectroscopic metallicities $\rm[Fe/H]_{L20}$ and our photometric $\rm[Fe/H]_{phot}$ estimates is shown in the left panel of Figure\,3. These estimates are quite consistent with one another, with an offset of $-0.05$\,dex and a much-improved scatter of 0.24\,dex (compared to the result of C22, see left panel of Figure\,2). 
We note that a mild bias of around 0.3\,dex is found at $\rm[Fe/H]_{phot}\sim -2.9$, although large weights are assigned for metal-poor sample stars as mentioned above.
This is a common issue for current calibrations, e.g., the efforts by \cite{2021ApJ...920...33D} and \cite{2022ApJS..261...33D}, and one can expect to reduce these systematics by assembly of more metal-poor RRLs with precise metallicity measurements from high-resolution spectroscopy in the near future.
 
Inspired by the studies of \cite{2021ApJ...920...33D} again, a first-order 2D polynomial fitting is applied to the [Fe/H] estimates for type RRc stars from L20, as a function of their period and $R_{21}$ from C22:
\begin{equation}
\begin{split}
{\rm[Fe/H]} = a_{0} + a_{1}(P-0.3)+ a_{2}(R_{21}-0.2) \text{,}
\end{split}
\end{equation}
where  $a_{i}$ ($i$ = 0, 1, 2) are the fit coefficients. Again, density-dependent sample weights and three-sigma clipping are performed in the fitting process. The fit coefficients are also listed in Table\,1. The right panel of Figure\,3 shows that the photometric metallicities resulted from Equation\,2 are in excellent agreement with those of the spectroscopic ones, with an offset of $-0.03$\,dex, and a small scatter of 0.19\,dex.

We caution that the above relations could be affected by the well-known Blazhko effect, a phenomenon of amplitude and/or phase modulation with a quasi-period of tens to a hundred days, first found by Blazhko (1907). The origin of this phenomenon remains uncertain. Measurements of $\phi_{31}$ and $R_{21}$ are affected for RRLs suffering a Blazhko effect, causing deviations of their $P$--$\phi_{31}$--$R_{21}$--[Fe/H] and $P$--$R_{21}$--[Fe/H] relations. 
As reported by previous studies (e.g., \citealt{2016pas..conf...22S,2018MNRAS.480.1229N}), the incidence rate of this effect for type RRab stars is  much greater than that for the type RRc stars. 
In addition, the uncertainty of the snapshot-derived spectroscopic metallicity (from L20) of type RRab stars is larger than that of type RRc stars, due to the larger amplitudes for the former. 
These may explain why the $P$--$R_{21}$--[Fe/H] relation of type RRc stars is much tighter than that of the type RRab stars, as well as the presence of fewer outliers in the fits for type RRc stars than for type RRab stars (red crosses in Figure\,3).

The newly constructed relations are then applied to those RRLs with precise measurements of period, $\phi_{31}$, and $R_{21}$ from C22 to derive their photometric metallicities. In total, estimates of metallicities for 135,873 RRLs (115,410 type RRab and 20,463 type RRc stars) are obtained for our final RRL sample. Uncertainties in the photometric metallicity estimates are primarily from two sources: errors in the period, $\phi_{31}$, and $R_{21}$ measurements and fit coefficient errors, and from the method itself. 
The former error can be estimated by a Monte Carlo (MC) simulation.
For each star, 1000 MC simulations are performed by sampling the measurement uncertainties of the period, $\phi_{31}$, and $R_{21}$ given by C22, and the errors of the fit coefficients in Table\,1. The random error of each star is then estimated from the distribution yielded by the MC simulations.
 The latter errors are provided by the scatter in the fits described above (see also the lower panels of Figure\,3).

The metallicity distribution of our final RRL sample is presented in Figure\,4, while the distribution of the mean metallicity for this sample in Galactic coordinates is shown in Figure\,5. 
This figure exhibits a negative gradient of metallicity from the Galactic plane toward toward high Galactic latitude, in agreement with our expectation.  Moreover, the Large Magellanic cloud (LMC) and Small Magellanic cloud (SMC) are clearly seen in the map.

\begin{table}
\centering
\caption{Fit Coefficients}
\begin{threeparttable}
\setlength{\tabcolsep}{5mm}{
\begin{tabular}{lcc}
\hline
\begin{minipage}{5mm}\vspace{4.5mm} \vspace{1mm} \end{minipage} 
Coeff.  & ${\rm[Fe/H]_{RRab}^a}$ & ${\rm[Fe/H]_{RRc}^b}$ \\ 
\hline
$a_{0}$ & $-1.888 \pm 0.002$ & $-1.737 \pm 0.005$\\
$a_{1}$ & $-5.772 \pm 0.026$ & $-9.968 \pm 0.079$\\
$a_{2}$ & $1.090 \pm 0.005$ & $-5.041 \pm 0.051$\\
$a_{3}$ & $1.065 \pm 0.030$ & $-$\\

\hline
\end{tabular}}
\begin{tablenotes}
\item[a] The fitting function is given by Equation\,(1).
\item[b] The fitting function is given by Equation\,(2).
\end{tablenotes}
\end{threeparttable}
\end{table}

\subsection{Distance}
The RRLs are well-behaved distance indicators, given the well-known absolute magnitude-metallicity relations (e.g., $M_{G}$--[Fe/H]) in the visual bands and period-absolute magnitude-metallicity relation ($PMZ$) in the infrared bands. Calibrations of these relations requires a large sample of RRLs with a wide metallicity distribution and accurate measurements of absolute magnitudes. With a large sample of photometric-metallicity estimates for RRLs, one can re-calibrate the relations by selecting local bright RRLs with both photometric-metallicity estimates given by this study and accurate distance measurements from the parallaxes reported in $Gaia$ EDR3. We define the calibration sample by cross-matching our final RRL sample to $Gaia$ EDR3 with the following cuts:
\begin{itemize}[leftmargin=*]
\item For type RRab stars, the value of $E(B-V)$ must be less than 0.1\,mag, either from \citet[][hereafter SFD98]{1998ApJ...500..525S} \footnote{The values of $E (B-V)$ from SFD98 are corrected for a systematic of 14\%, as reported in previous studies (e.g., \citealt{2010ApJ...725.1175S,2013MNRAS.430.2188Y})} for high-latitude regions ($|b|\ge 25^{\circ}$), or from C22 for low-latitude regions ($|b| < 25^{\circ}$);

\item For type RRc stars, the value of $E(B-V)$ must be less than 0.2\,mag, given by SFD98 for high-latitude regions with $|b|\ge 20^{\circ}$;

\item The RRLs must have parallaxes greater than 0.25\,mas, and relative parallaxes error smaller than 10 per cent;

\item The error of photometric metallicities must be less than 0.3 and 0.2\,dex for type RRab and RRc stars, respectively;

\item The RRLs must have \emph{phot\_bp\_rp\_excess\_factor} $<$1.5 and \emph{astrometric excess noise} $<$0.25 given by $Gaia$ EDR3.

\end{itemize}

The first three cuts ensure precise determinations of the $G$, $K\rm_{s}$-band, and $W1$-band absolute magnitudes. The fourth cut is to ensure reliable photometric metallicity estimates. The last cut is to exclude potential stellar binaries, blends, and galaxies. In the second cut, no RRc stars in low-latitude regions are selected, due to the lack of accurate extinction estimates in the disk region. To ensure a sufficient number of RRc stars for the calibration, the definition of the high-latitude region is slightly looser for RRc stars than that for type RRab stars. Finally, a total of 825 type RRab and 100 type RRc stars are selected for the calibrations (hereafter the RRL\_CAL\_DIS sample). 

The distances of the selected RRLs are derived from the parallax measurements from $Gaia$ EDR3 by a Bayesian method, as that used by \cite{2021ApJ...907L..42H}. We note that the zero-points of $Gaia$ EDR3 parallax have been corrected using the procedures provided by \cite{2021A&A...649A...4L}, which has been independently examined by other studies (e.g., \citealt{2021ApJ...910L...5H,2021ApJ...911L..20R,2021AJ....161..214Z,2022AJ....163..149W}).

\begin{figure}[!htbp]
\begin{minipage}[t]{0.5\linewidth}
\centering
\includegraphics[width=3.4in]{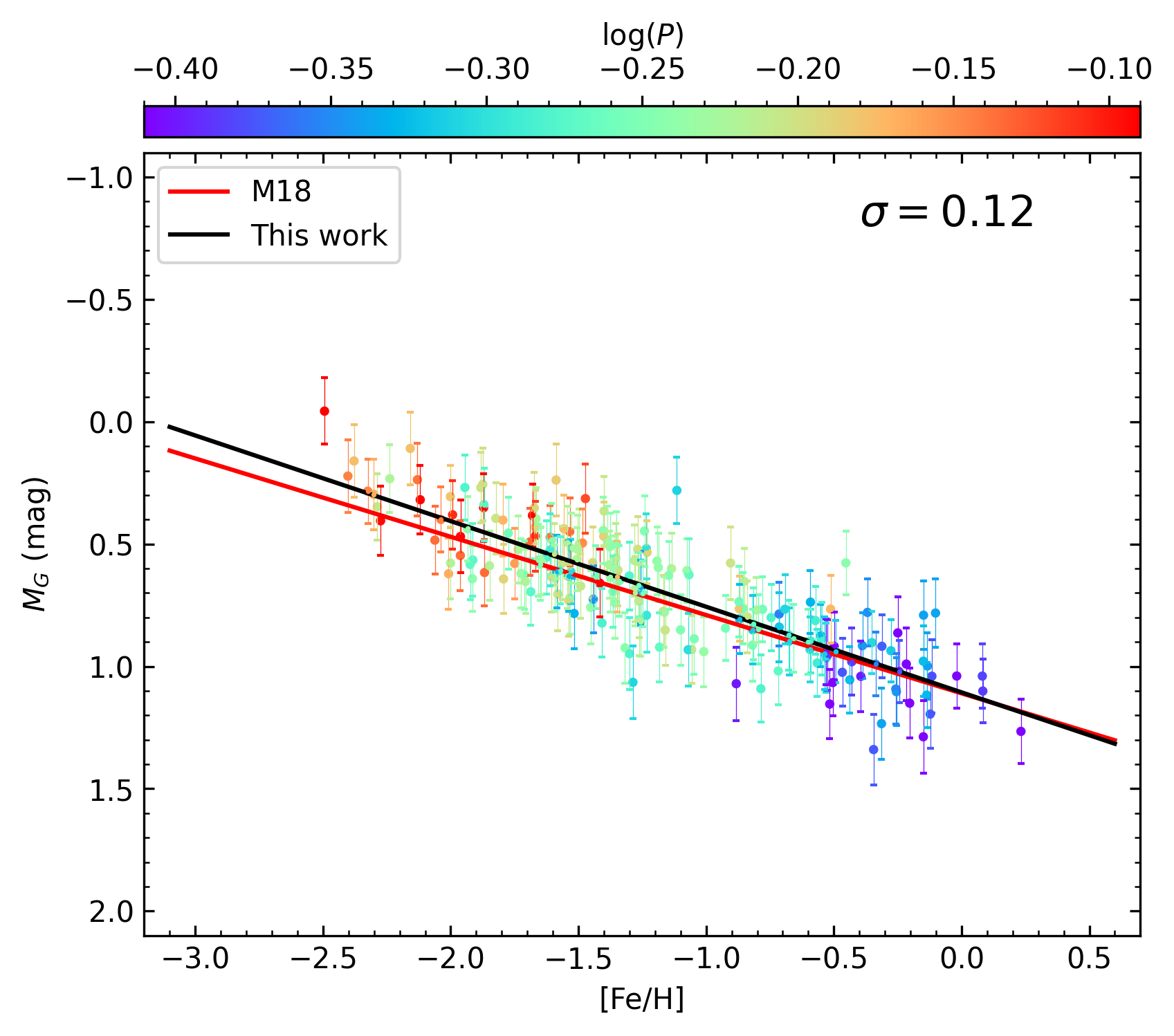}
\end{minipage}
\caption{$M_{G}$--[Fe/H] distribution of 205 type RRab stars from our RRL\_CAL\_DIS sample, color-coded by the pulsation period on a logarithmic scale, as shown in the color bar. The black line represents the best-fit result. The red line is the fit result of Muraveva et al. (2018). The standard deviation of fitting residuals is marked in the top-right corner.}
\end{figure}

\begin{figure}[!htbp]
\begin{minipage}[t]{0.5\linewidth}
\centering
\includegraphics[width=3.35in]{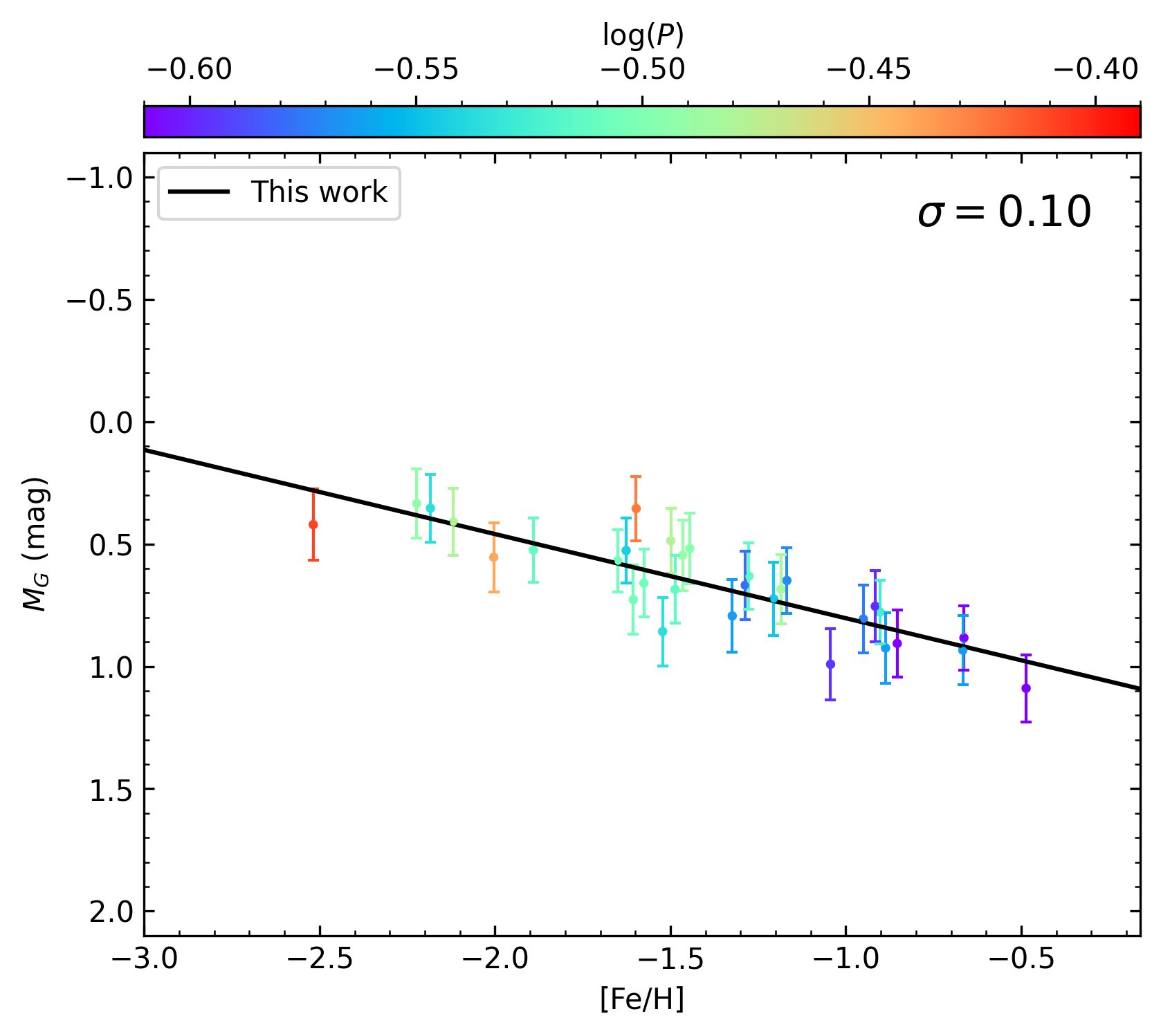}
\end{minipage}
\caption{Same as Figure\,6, but for RRc stars.}
\end{figure}

\begin{figure}[!htbp]
\begin{minipage}[t]{0.5\linewidth}
\centering
\includegraphics[width=3.35in]{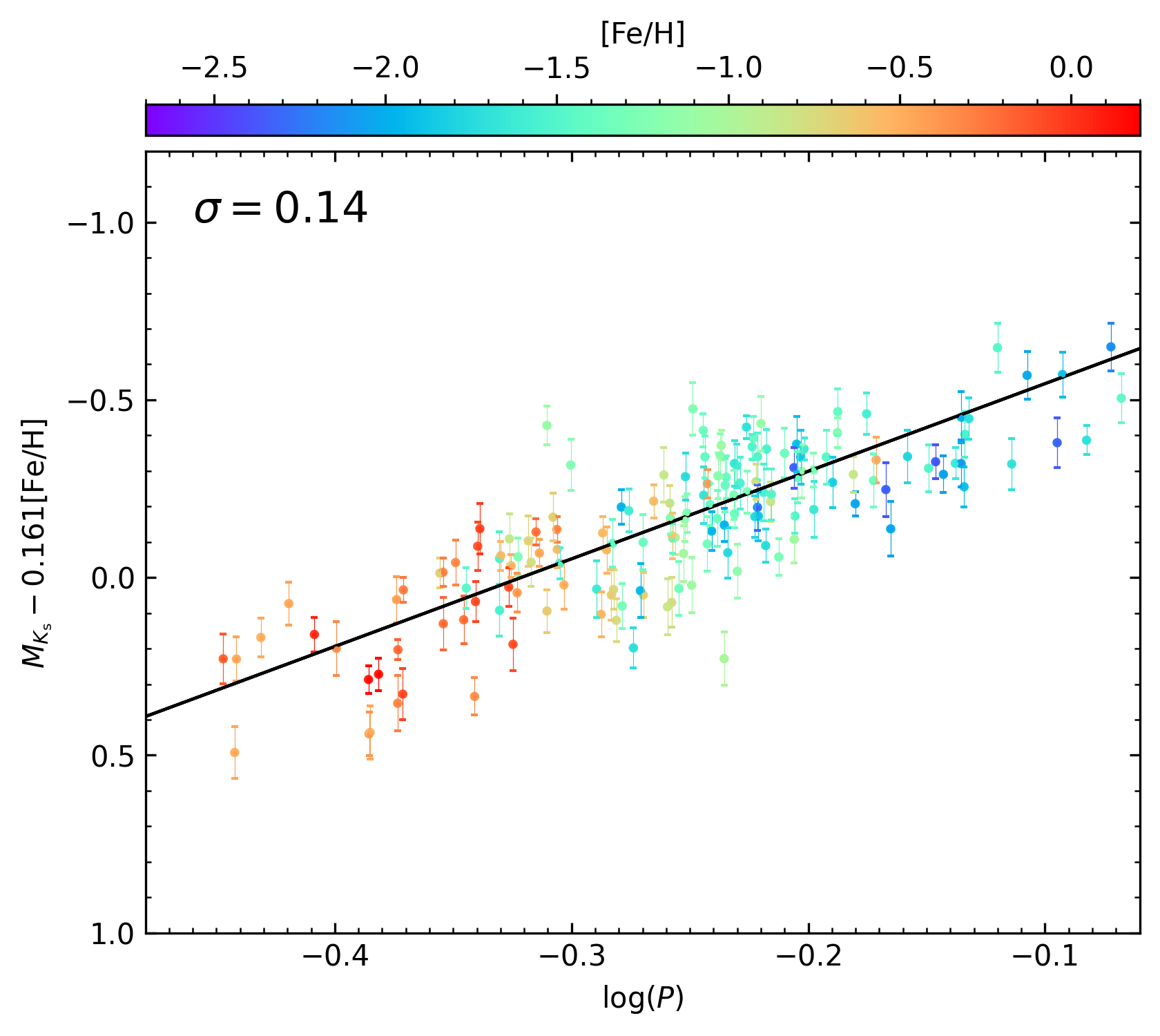}
\end{minipage}
\caption{$PM_{K{\rm_{s}}}Z$ relation for 159 type RRab stars, color-coded by metallicity. as shown in the color bar. The black line represents the best-fit result. The standard deviation of fitting residuals is marked in the top-left corner.}
\end{figure}

\begin{figure}[!htbp]
\begin{minipage}[t]{0.5\linewidth}
\centering
\includegraphics[width=3.35in]{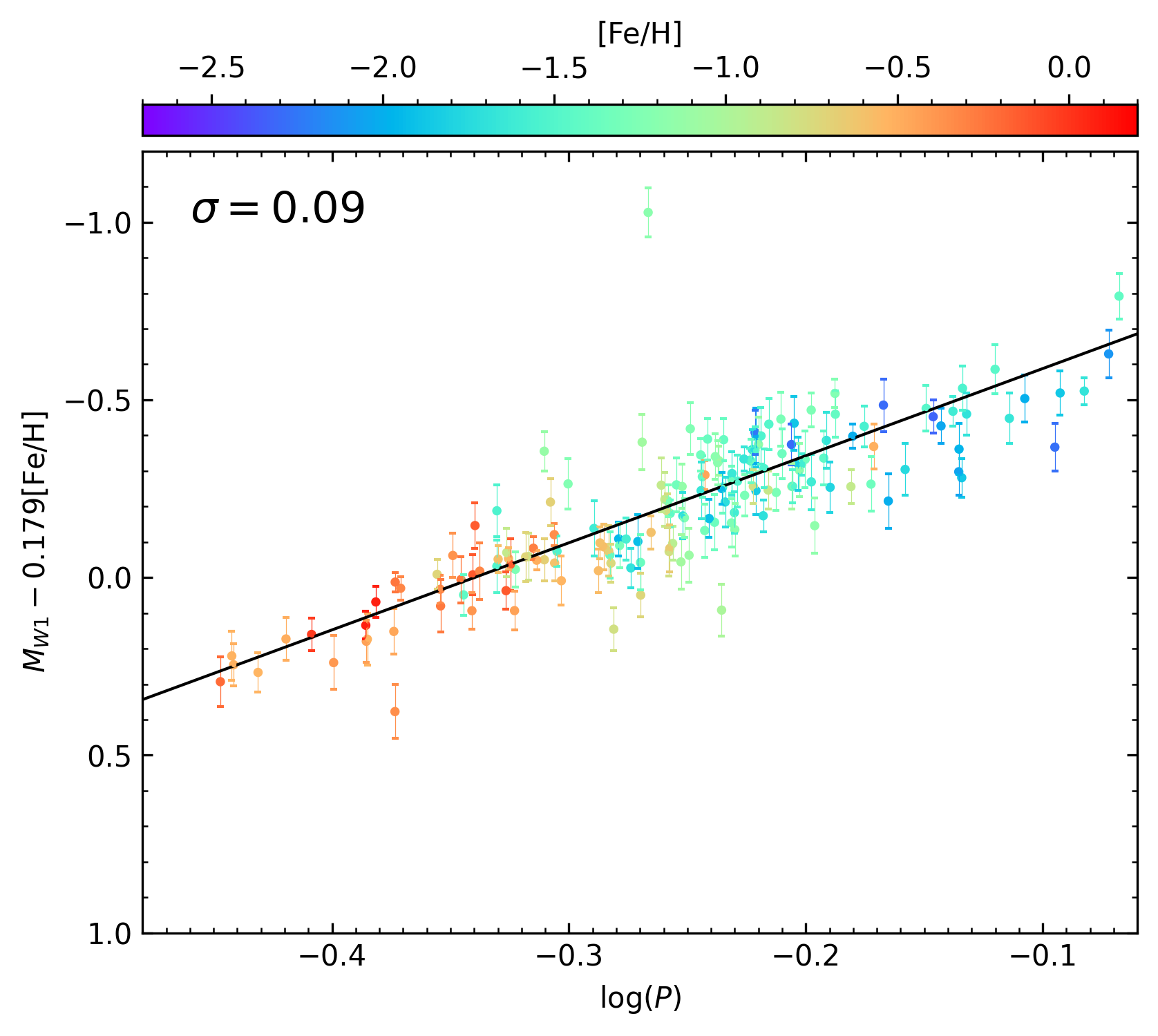}
\end{minipage}
\caption{Same as Figure\,8, but for the $PM_{W1}Z$ relation of 164 type RRab stars.}
\end{figure}

\begin{figure*}
\includegraphics[width=2.3in]{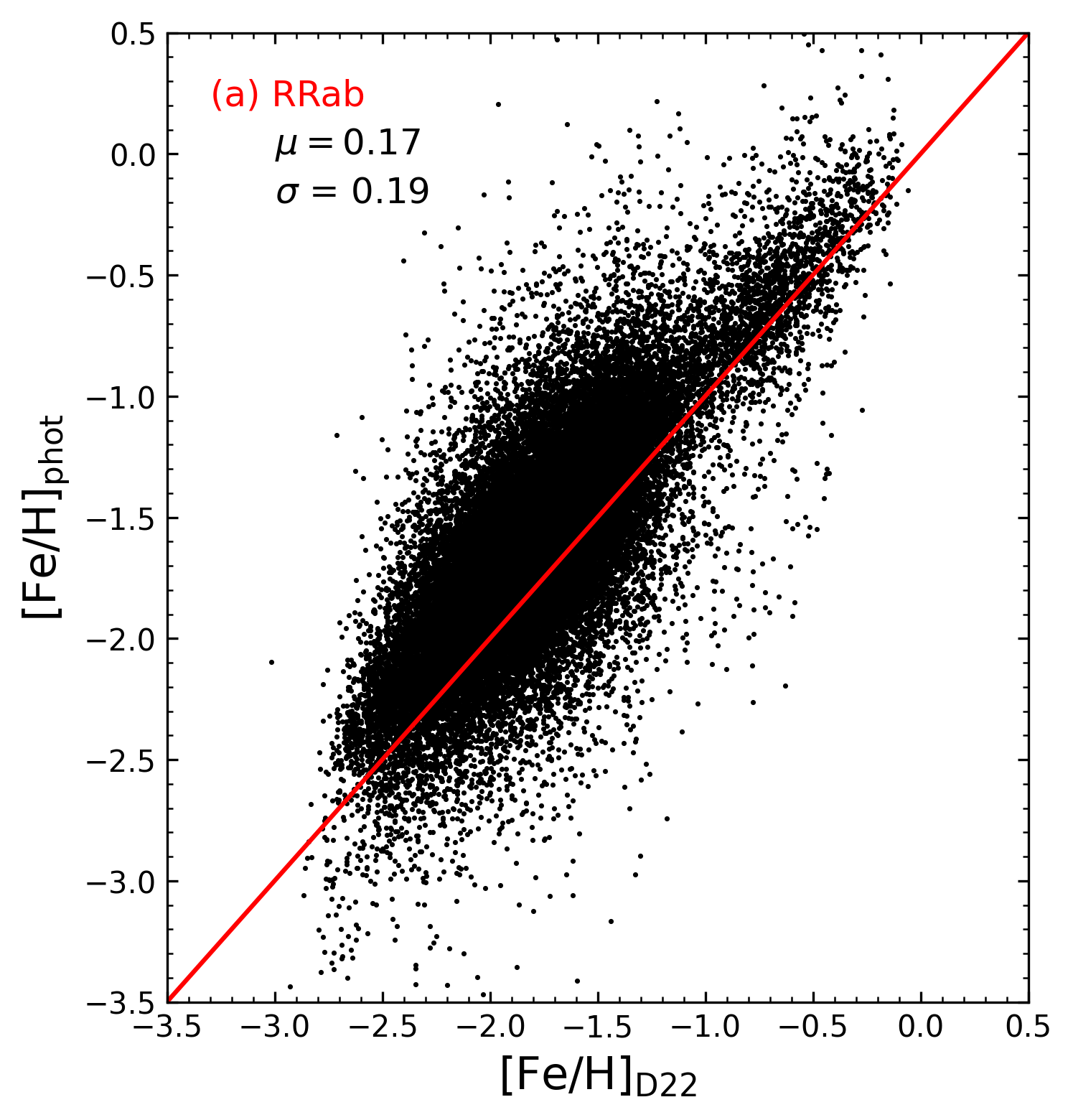}
\includegraphics[width=2.3in]{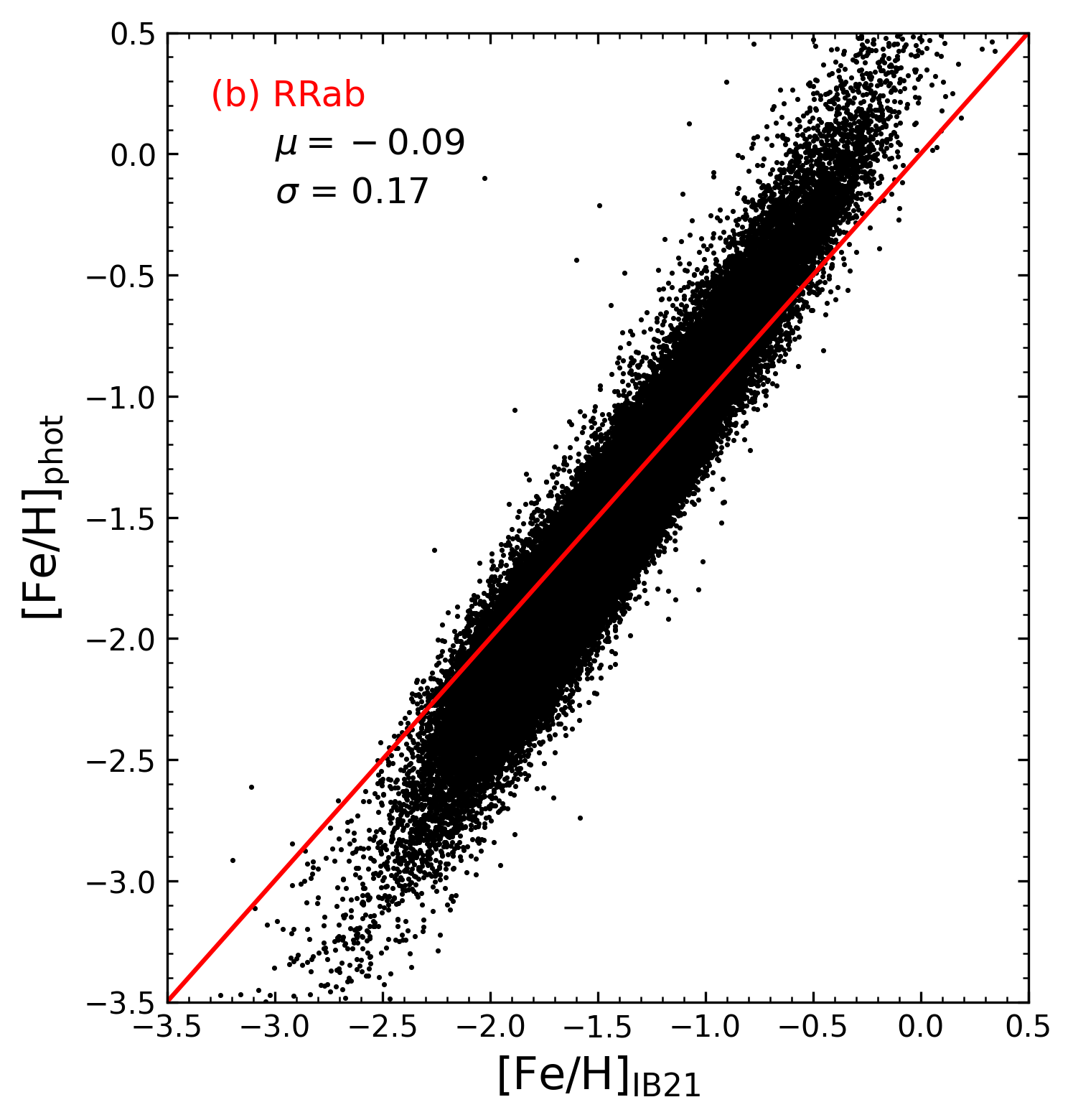}
\includegraphics[width=2.3in]{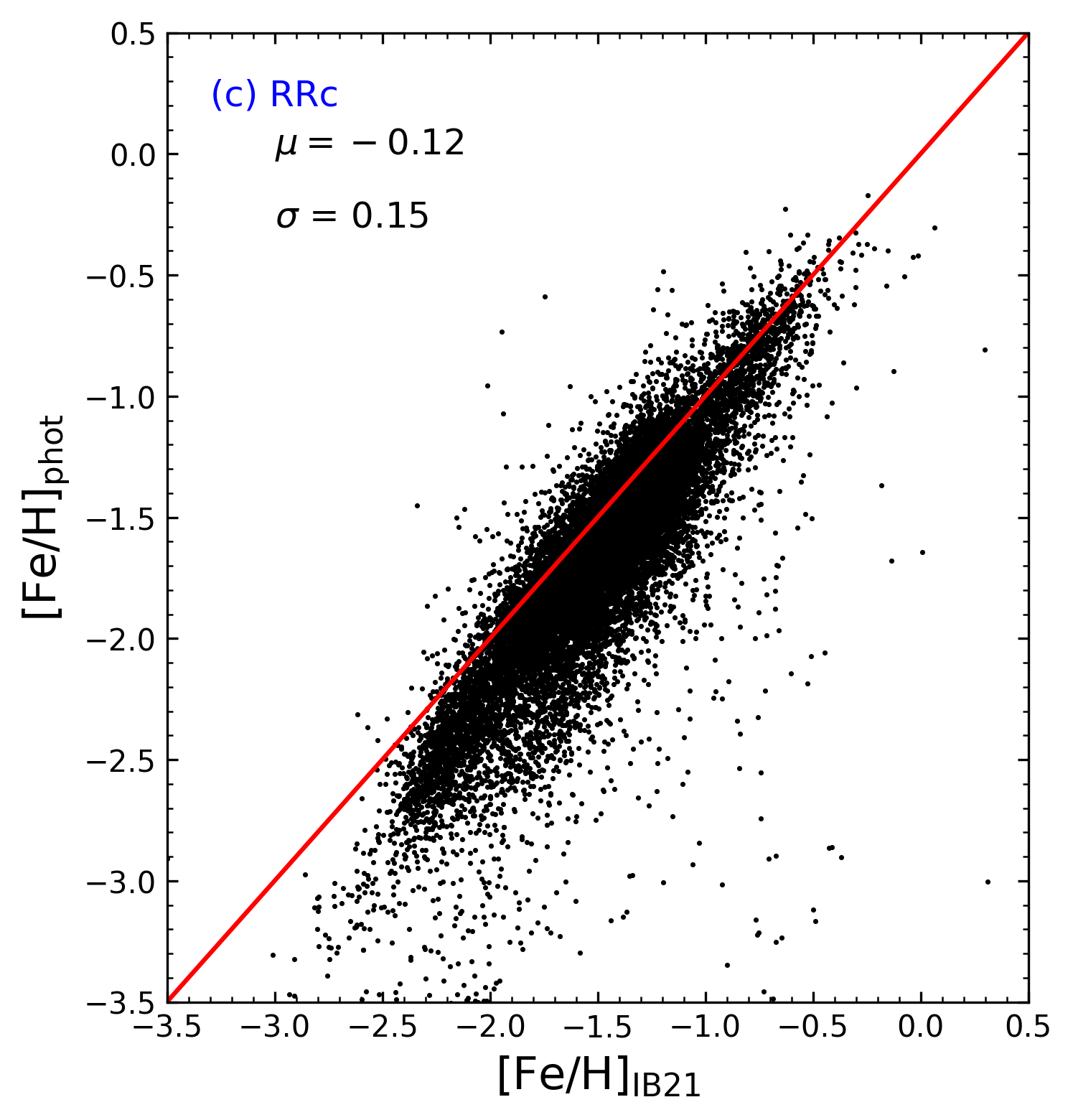}
\caption{{\it Panel (a):} Comparison of metallicity for type RRab stars from this work and those measured by D22. {\it Panel (b):} Similar to panel (a) but for the comparison between this work and IB21. {\it Panel (c):} Similar to panel (b) but for type RRc stars. The values of mean and standard deviation of the metallicity differences (this work minus D22 or IB21) are marked in the top-left corner of each panel.}
\end{figure*}

\begin{figure*}
\begin{minipage}[t]{0.5\linewidth}
\centering
\includegraphics[width=6.8in]{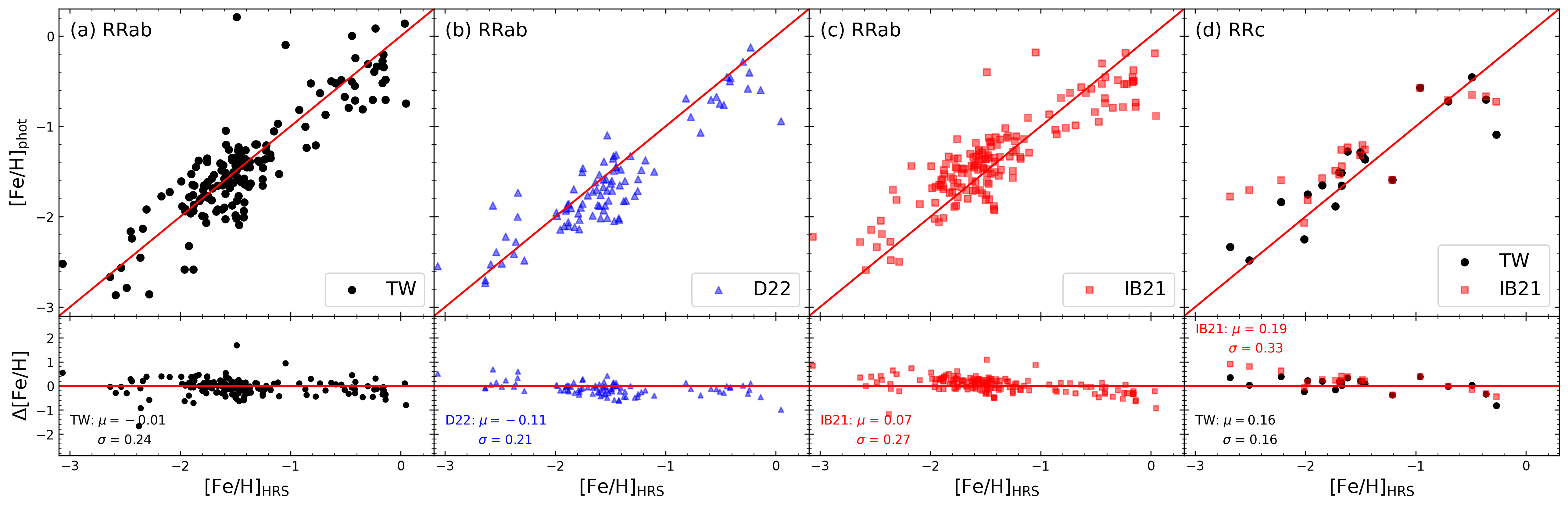}
\end{minipage}
\caption{{\it Panel (a)}: Comparison of metallicity for type RRab stars from this work (TW) and those measured from  high-resolution spectroscopy \citep[HRS;][]{2021ApJ...920...33D}.
{\it Panel (b):} Similar to panel (a) but for comparison between D22 and HRS.
{\it Panel (c):} Similar to panel (a) but for comparison between IB21 and HRS.
{\it Panel (d):} Similar to panel (a) but for comparison for type RRc stars between TW (black dots), IB21 (red squares) and HRS.
The metallicity differences {$\rm\Delta[Fe/H]$} (TW/D22/IB21 minus HRS) are shown in the lower part of each panel, with the mean and standard deviation marked in the bottom-left corner.}
\end{figure*}

\begin{table*}
\centering
\caption{$M_{G}$--[Fe/H] and $PMZ$ Relations from the RRL\_CAL\_DIS Sample}
\begin{threeparttable}
\setlength{\tabcolsep}{4.5mm}{
\begin{tabular}{ccccc}
\hline
\begin{minipage}{5mm}\vspace{4.mm} \vspace{1mm} \end{minipage} 
Relation & Mathematical form & $M$\,(mag) & $N$ &$\sigma$\,(mag)\\ 
\hline
$M_{G}$--[Fe/H]\tnote{a} & $M_{G}=(0.350 \pm 0.016)\,{\rm[Fe/H]}+(1.106 \pm 0.021)$ &$0.584 \pm 0.036$\tnote{c} & $205$ & 0.12\\
$M_{G}$--[Fe/H]\tnote{b} & $M_{G}=(0.344 \pm 0.051)\,{\rm[Fe/H]}+(1.147 \pm 0.074)$ &$0.631 \pm 0.103$\tnote{c} & $31$ & 0.10\\
$PM_{K_{\rm {s}}}Z$\tnote{a} & $M_{K_{\rm {s}}}=(-2.465 \pm 0.084)\,{\rm log}(P)+(0.161 \pm 0.011)\,{\rm[Fe/H]}-(0.792 \pm 0.043)$ &$-0.341 \pm 0.043$\tnote{d} & $159$ & 0.14\\
$PM_{W1}Z$\tnote{a} &  $M_{W1}=(-2.452 \pm 0.080)\,{\rm log}(P)+(0.179 \pm 0.011)\,{\rm[Fe/H]}-(0.834 \pm 0.031)$ &$-0.413 \pm 0.040$\tnote{d} & $164$ & 0.09\\

\hline
\end{tabular}}
\begin{tablenotes}
\item[a] The fitting result is given for type RRab stars.
\item[b] The fitting result is given for type RRc stars.
\item[c] Absolute magnitudes of type RRab or type RRc stars in different passbands, calculated by adopting a metallicity [Fe/H] = $-$1.5.
\item[d] Absolute magnitudes of type RRab stars in different passbands, calculated by adopting a metallicity [Fe/H] = $-$1.5 and $P$ = 0.5238\,days.
\end{tablenotes}
\end{threeparttable}
\end{table*}

\begin{table}
\centering
\caption{Summary of $E(B - V)$ Determinations}
\begin{threeparttable}
\setlength{\tabcolsep}{3.8mm}{
\begin{tabular}{lccc}
\hline
\begin{minipage}{5mm}\vspace{4.5mm} \vspace{1mm} \end{minipage} 
Region  & Method & Flag & $N$ \\ 
\hline
All sky & SFD98 & sfd98 & 31074\tnote{a} /19097\tnote{b}\\
$|b| < 25^{\circ}$ & C22   & c22   & 64364\tnote{a} /0\tnote{b}\\
Magellanic Clouds & S21   & s21   & 19541\tnote{a} /1366\tnote{b}\\
Globular clusters  & H10   & h10   & 431\tnote{a} /78\tnote{b}\\
 
\hline
\end{tabular}}
\begin{tablenotes}
\item[a] The number of type RRab stars.
\item[b] The number of type RRc stars.
\end{tablenotes}
\end{threeparttable}
\end{table}

\begin{table*}
\centering
\caption{ Comparison of Photometric Metallicities and Distances for GCs with Values from H10 and B19}
\begin{threeparttable}
\begin{tabular}{ccccccccccc}

\hline\hline
Name&[Fe/H]$_{\rm H10}$&$d_{\rm H10}$&$d_{\rm B19}$&$<{\rm [Fe/H]}_{\rm C22}>$&$\sigma_{\rm [Fe/H]_{\rm C22}}$&$<{\rm [Fe/H]}_{\rm phot}>$&$\sigma_{\rm [Fe/H]_{\rm phot}}$&$<d_{\rm phot}>$&$\sigma_{d_{\rm phot}}$&$N$\\
&&(kpc)&(kpc)&(dex)&(dex)&(dex)&(dex)&(kpc)&(kpc)&\\

\hline
 \noalign{\smallskip}
\multicolumn{11}{c}{(a) RRab} \\ \hline
 \noalign{\smallskip}
NGC 1851&$-1.18$&$12.1$&$11.32\pm0.20$&$-1.23$&$0.54$&$-1.60$&$0.43$&$12.04$&$0.91$&5\\
 NGC 3201&$-1.59$&$4.9$&$4.47\pm0.18$&$-1.23$&$0.38$&$-1.58$&$0.19$&$4.78$&$0.24$&41\\
 Rup 106&$-1.68$&$21.2$&$-$&$-1.26$&$0.24$&$-1.77$&$0.11$&$21.20$&$0.42$&8\\
 NGC 4590&$-2.23$&$10.3$&$-$&$-1.87$&$0.66$&$-2.10$&$0.29$&$10.16$&$0.59$&8\\
 NGC 4833&$-1.85$&$6.6$&$6.10\pm0.43$&$-1.55$&$0.34$&$-1.96$&$0.13$&$6.21$&$0.34$&5\\
 NGC 5024&$-2.10$&$17.9$&$-$&$-1.56$&$0.62$&$-2.00$&$0.18$&$17.88$&$0.58$&17\\
 NGC 5053&$-2.27$&$17.4$&$-$&$-1.68$&$0.28$&$-2.07$&$0.13$&$16.70$&$0.16$&5\\
 NGC 5272&$-1.50$&$10.2$&$9.47\pm0.45$&$-1.38$&$0.43$&$-1.69$&$0.20$&$10.15$&$0.39$&74\\
 NGC 5466&$-1.98$&$16.0$&$-$&$-1.70$&$0.31$&$-2.01$&$0.21$&$15.71$&$0.33$&11\\
 IC 4499&$-1.53$&$18.8$&$-$&$-1.40$&$0.48$&$-1.73$&$0.28$&$19.35$&$1.00$&46\\
 NGC 5824&$-1.91$&$32.1$&$-$&$-1.26$&$0.77$&$-1.84$&$0.25$&$30.83$&$1.48$&5\\
 NGC 5904&$-1.29$&$7.5$&$7.58\pm0.14$&$-1.28$&$0.48$&$-1.60$&$0.27$&$7.35$&$0.37$&14\\
 NGC 6121&$-1.16$&$2.2$&$1.96\pm0.04$&$-1.00$&$0.16$&$-1.39$&$0.12$&$2.05$&$0.17$&5\\
 NGC 6171&$-1.02$&$6.4$&$5.92\pm0.38$&$-0.58$&$0.36$&$-1.14$&$0.12$&$5.88$&$0.37$&7\\
 NGC 6229&$-1.47$&$30.5$&$-$&$-1.04$&$0.31$&$-1.47$&$0.17$&$29.38$&$0.94$&11\\
 NGC 6266&$-1.18$&$6.8$&$6.40\pm0.18$&$-0.72$&$0.37$&$-1.30$&$0.26$&$6.15$&$0.67$&11\\
 NGC 6362&$-0.99$&$7.6$&$7.34\pm0.31$&$-0.79$&$0.34$&$-1.29$&$0.25$&$7.26$&$0.26$&13\\
 NGC 6401&$-1.02$&$10.6$&$-$&$-0.63$&$0.28$&$-1.11$&$0.18$&$8.80$&$0.63$&7\\
 NGC 6402&$-1.28$&$9.3$&$9.31\pm0.50$&$-0.73$&$0.33$&$-1.27$&$0.22$&$8.60$&$0.44$&14\\
 NGC 6426&$-2.15$&$20.6$&$-$&$-1.72$&$0.43$&$-2.15$&$0.10$&$20.73$&$0.45$&8\\
 NGC 6584&$-1.50$&$13.5$&$-$&$-1.08$&$0.38$&$-1.54$&$0.26$&$13.16$&$0.55$&24\\
 NGC 6715&$-1.49$&$26.5$&$24.15\pm0.38$&$-1.26$&$0.75$&$-1.57$&$0.42$&$26.94$&$1.76$&15\\
 NGC 6934&$-1.47$&$15.6$&$14.57\pm1.43$&$-1.33$&$0.42$&$-1.67$&$0.22$&$15.60$&$0.54$&29\\
 NGC 6981&$-1.42$&$17.0$&$-$&$-1.38$&$0.31$&$-1.70$&$0.13$&$16.84$&$0.45$&24\\
 NGC 7006&$-1.52$&$41.2$&$-$&$-1.35$&$0.51$&$-1.65$&$0.24$&$42.11$&$1.69$&11\\
 NGC 7078&$-2.37$&$10.4$&$10.21\pm0.13$&$-2.01$&$0.45$&$-2.21$&$0.17$&$10.69$&$0.65$&13\\

\hline
 \noalign{\smallskip}

\multicolumn{11}{c}{(b) RRc} \\ \hline
 \noalign{\smallskip}
NGC 1851&$-1.18$&$12.1$&$11.32\pm0.20$&$-1.10$&$0.10$&$-1.53$&$0.04$&$11.97$&$0.27$&3\\
 NGC 4590&$-2.23$&$10.3$&$-$&$-2.20$&$0.06$&$-2.35$&$0.15$&$10.60$&$0.23$&8\\
 NGC 5053&$-2.27$&$17.4$&$-$&$-1.96$&$0.33$&$-2.21$&$0.15$&$17.33$&$0.72$&4\\
 NGC 5272&$-1.50$&$10.2$&$9.47\pm0.45$&$-1.52$&$0.13$&$-1.56$&$0.24$&$9.71$&$0.41$&9\\
 NGC 5466&$-1.98$&$16.0$&$-$&$-1.77$&$0.30$&$-1.72$&$0.23$&$15.27$&$0.50$&3\\
 IC 4499&$-1.53$&$18.8$&$-$&$-1.79$&$0.17$&$-1.84$&$0.09$&$19.28$&$0.22$&6\\
 NGC 6121&$-1.16$&$2.2$&$1.96\pm0.04$&$-0.98$&$0.13$&$-1.29$&$0.10$&$2.07$&$0.10$&3\\
 NGC 6171&$-1.02$&$6.4$&$5.92\pm0.38$&$-0.97$&$0.08$&$-1.19$&$0.10$&$5.76$&$0.15$&6\\
 NGC 6266&$-1.18$&$6.8$&$6.40\pm0.18$&$-1.12$&$0.06$&$-1.28$&$0.09$&$6.48$&$0.59$&3\\
 NGC 6362&$-0.99$&$7.6$&$7.34\pm0.31$&$-0.95$&$0.13$&$-1.22$&$0.16$&$7.19$&$0.19$&8\\
 NGC 6402&$-1.28$&$9.3$&$-$&$-0.87$&$0.29$&$-1.29$&$0.18$&$8.93$&$0.19$&5\\
 NGC 6584&$-1.50$&$13.5$&$-$&$-1.50$&$0.15$&$-1.62$&$0.11$&$13.31$&$0.19$&4\\
 NGC 6638&$-0.95$&$9.4$&$-$&$-1.00$&$0.11$&$-1.28$&$0.04$&$9.65$&$0.18$&3\\
 NGC 7078&$-2.37$&$10.4$&$10.21\pm0.13$&$-2.16$&$0.28$&$-2.47$&$0.09$&$11.06$&$0.19$&13\\
 
\hline
\end{tabular}
\begin{tablenotes}
\item[] Col.\,1 gives the cluster identification number, Cols.\,2 and 3 give the metallicities and distances of the GCs from H10, Col.\,4 gives the distances and their errors of the GCs from B19, Cols. 5--10 present the mean values and their errors in C22, our photometric metallicities and distances of the GCs (see Section\,4.2) , and Col.\,11 gives the number of GC member RRL stars that pass the cuts presented in Section\,4.2.
\end{tablenotes}
\end{threeparttable}
\end{table*}

\begin{table*}
\centering
\caption{Description of the Final Sample}
\begin{threeparttable}
\setlength{\tabcolsep}{6mm}{
\begin{tabular}{llc}
\hline

Field  & Description & Unit \\ 
\hline
source\_id & Unique source identifier for $Gaia$ DR3 & ...\\
R.A.& Right ascension from $Gaia$ DR3 (J2016) & degrees\\
Decl. & Declination from $Gaia$ DR3 (J2016) & degrees\\
${\rm[Fe/H]}$\_PHOT & Photometric metallicity & dex\\
${\rm[Fe/H]}$\_PHOT\_ERROR & Uncertainty of photometric metallicity & dex\\
MG&     $G$-band absolute magnitude    & mag\\
MG\_ERROR&    Uncertainty of  $G$-band absolute magnitude          & mag\\
EBV\_ADOP&   Value of $E(B-V)$           & mag\\
EBV\_ADOP\_FLAG&   Flag to indicate the source of $E(B-V)$, which takes values ``sfd98",``c22", ``s21" and ``h10"    & ...\\
DIST\_PHOT&     Distance            & kpc\\
DIST\_PHOT\_ERROR&     Uncertainty of Distance            & kpc\\
Type&    Type of RR Lyrae star, which take values ``RRab" and ``RRc"   & ...\\
\hline
\end{tabular}}
\begin{tablenotes}
\item[] The final RRL sample with items defined above is provided online, as well as at \url{http://doi.org/10.5281/zenodo.6731860}.
\end{tablenotes}
\end{threeparttable}
\end{table*}

\begin{table}
\centering
\caption{Summary of the Error Budget for Distance and Metallicity Estimates of the LMC and SMC}
\begin{threeparttable}
\begin{tabular}{lcll}

\hline\hline
Description&Term&LMC&SMC\\

\hline
 \noalign{\smallskip}
\multicolumn{4}{c}{Distance} \\ \hline
 \noalign{\smallskip}
 \multicolumn{4}{c}{Statistical uncertainties (per star)} \\ 
 
Mean $G$-band magnitude uncertainty&$\sigma_{ G}$&$0.014$&$0.013$\\
Extinction\tnote{a}&$\sigma_{A_{ G}}$&$0.103$&$0.069$\\
$G$-band absolute magnitude&$\sigma_{M_{ G}}$&$0.134$&$0.137$\\
Total statistical uncertainty&--&0.001&0.003\\
 \multicolumn{4}{c}{Systematic uncertainties}\\
 {\it Gaia} photometric calibration\tnote{b}&$\sigma_{{\rm ZP}_{ G}}$&0.004&0.004\\
 Extinction\tnote{a}&$\sigma_{A_{ G},sys}$&$0.020$&$0.020$\\
 $M_G$--[Fe/H] relation metallicity constant\tnote{c}&$x\sigma_{b}$&0.027&0.032\\
 $M_G$--[Fe/H] relation zero-point&$\sigma_{c}$&0.021&0.021\\
 Total systematic uncertainty&--&0.040&0.043\\

\hline
 \noalign{\smallskip}

\multicolumn{4}{c}{Metallicity} \\ \hline
 \noalign{\smallskip}
 \multicolumn{4}{c}{Statistical uncertainties (per star)} \\ 
Metallicity&$\sigma_{\rm [Fe/H]}$&$0.361$&0.371\\
Total statistical uncertainty&--&0.003&0.008\\
\multicolumn{4}{c}{Systematic uncertainties}\\
$P$--$\phi_{31}$--[Fe/H] relation fit coefficients\tnote{d}&$\sigma_{a_{0}}$&0.002&0.002\\
&$y\sigma_{a_{1}}$&0.015&0.016\\
&$z\sigma_{a_{2}}$&0.010&0.010\\
&$t\sigma_{a_{3}}$&0.014&0.013\\
 Total systematic uncertainty&--&0.023&0.023\\

\hline
\end{tabular}
\begin{tablenotes}
\item[a] The typical uncertainties of the extinction and extinction scale are from \citep{2021ApJS..252...23S}.
\item[b] The typical calibration uncertainty of the {\it Gaia} $G$-band is taken from \citep{2021ApJ...908L..24Y}.
\item[c] Here $x$ represents the mean metallicity of LMC/SMC, given by in Section\,5. 
\item[d] Here $y$, $z$, and $t$ represent the mean period, $\phi_{31}$, and $R_{21}$ of type ab RRLs in LMC/SMC, taking values of 0.581/0.598\,days, 2.087/2.095 and 0.461/0.432, respectively. 
\end{tablenotes}
\end{threeparttable}
\end{table}

\begin{figure*}[!htbp]
\includegraphics[width=3.5in]{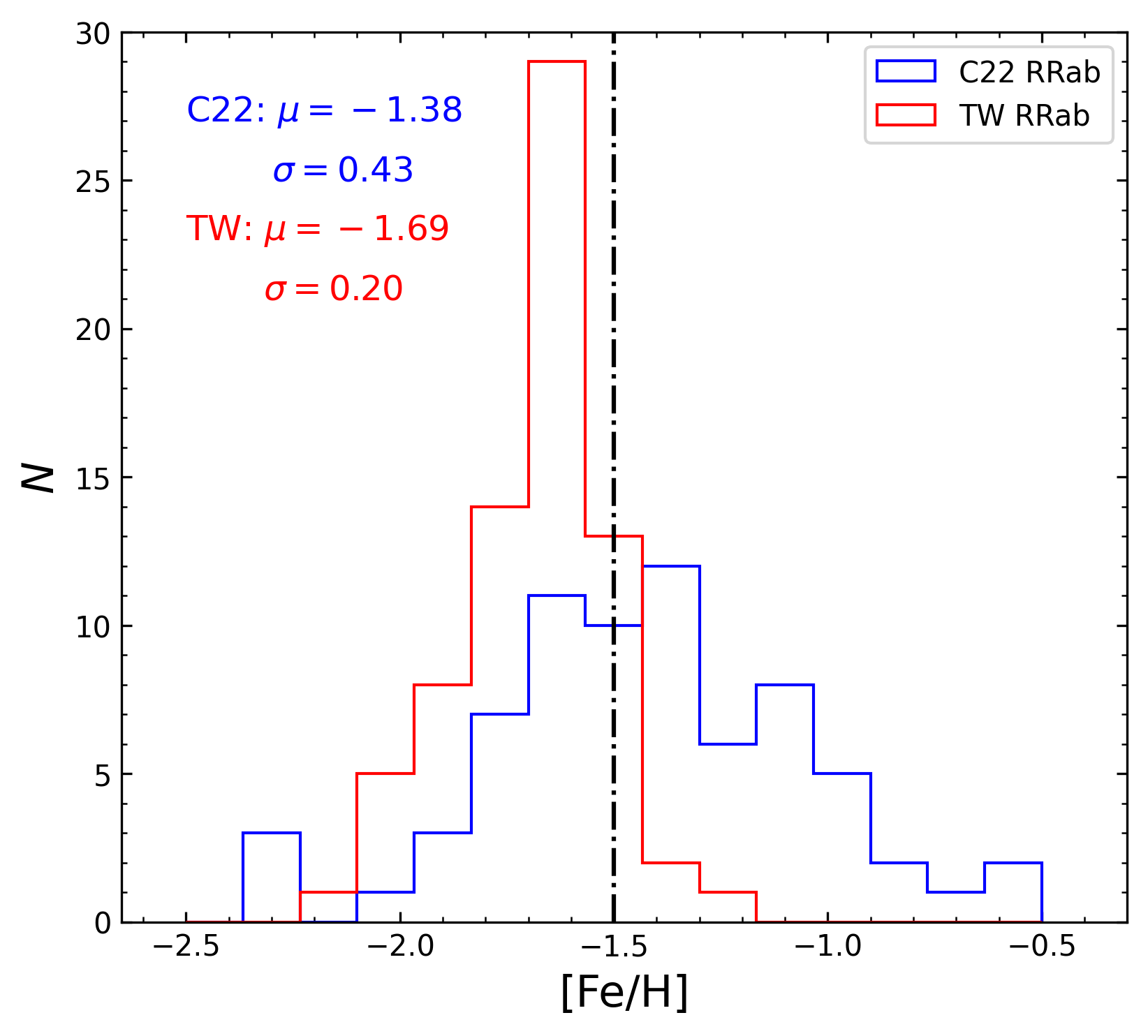}
\includegraphics[width=3.55in]{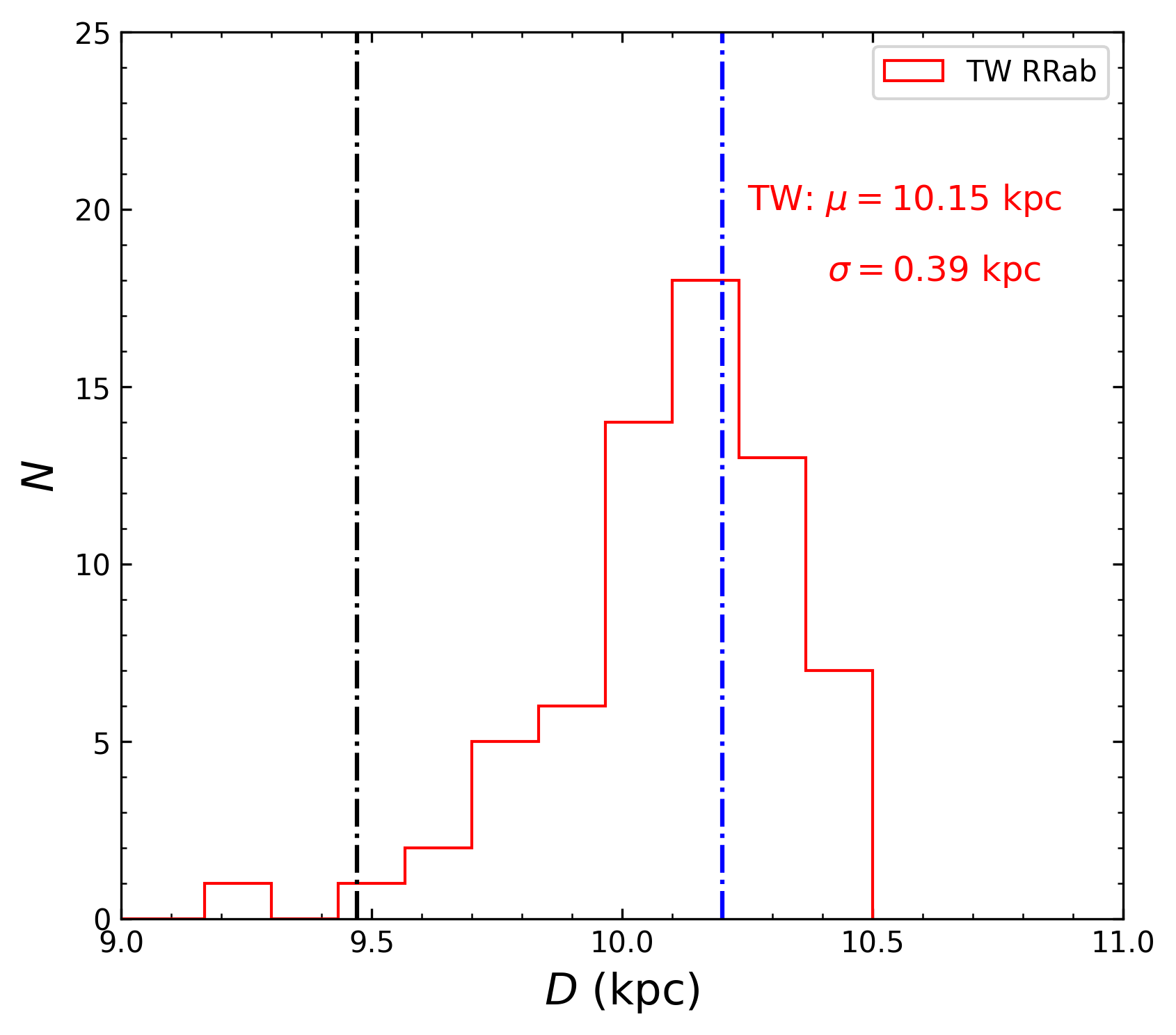}
\caption{{\it Left panel:} Metallicity distribution for member type RRab stars of NGC\,5272, with red for this work (TW) and blue for C22. The metallicity of NGC\,5272 from H10 is marked with the dash-dot line. The dispersion of the metallicity of TW is much narrower than that of C22. The means and standard deviations of the distributions are marked in the top-left corner.
{\it Right panel:} Similar to the left panel but for the distance distribution. The distance of NGC\,5272 from B19 and H10 are marked with the black and blue dash-dot lines, respectively.}
\end{figure*}

\begin{figure*}[!htbp]
\includegraphics[width=3.5in]{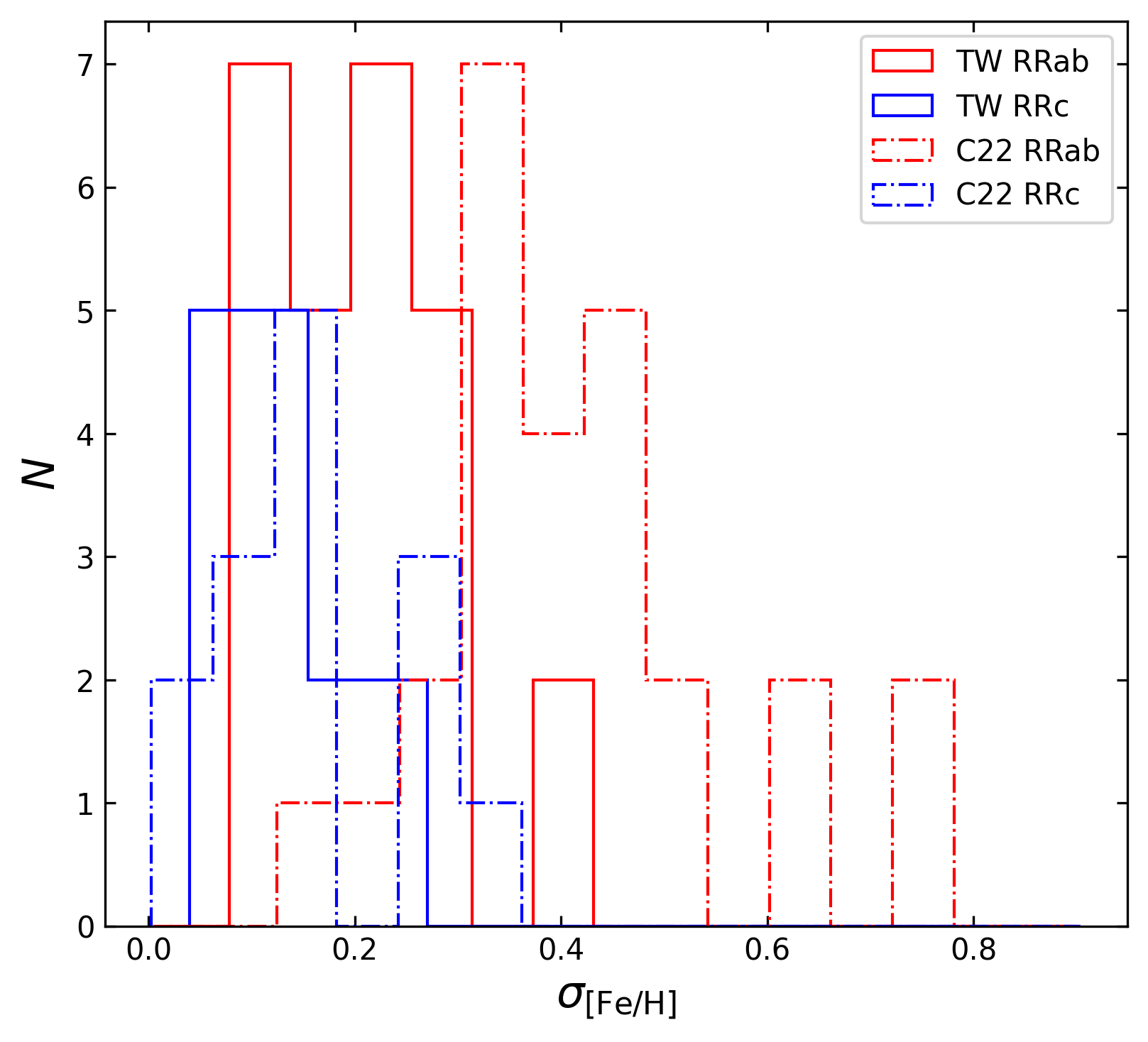}
\includegraphics[width=3.45in]{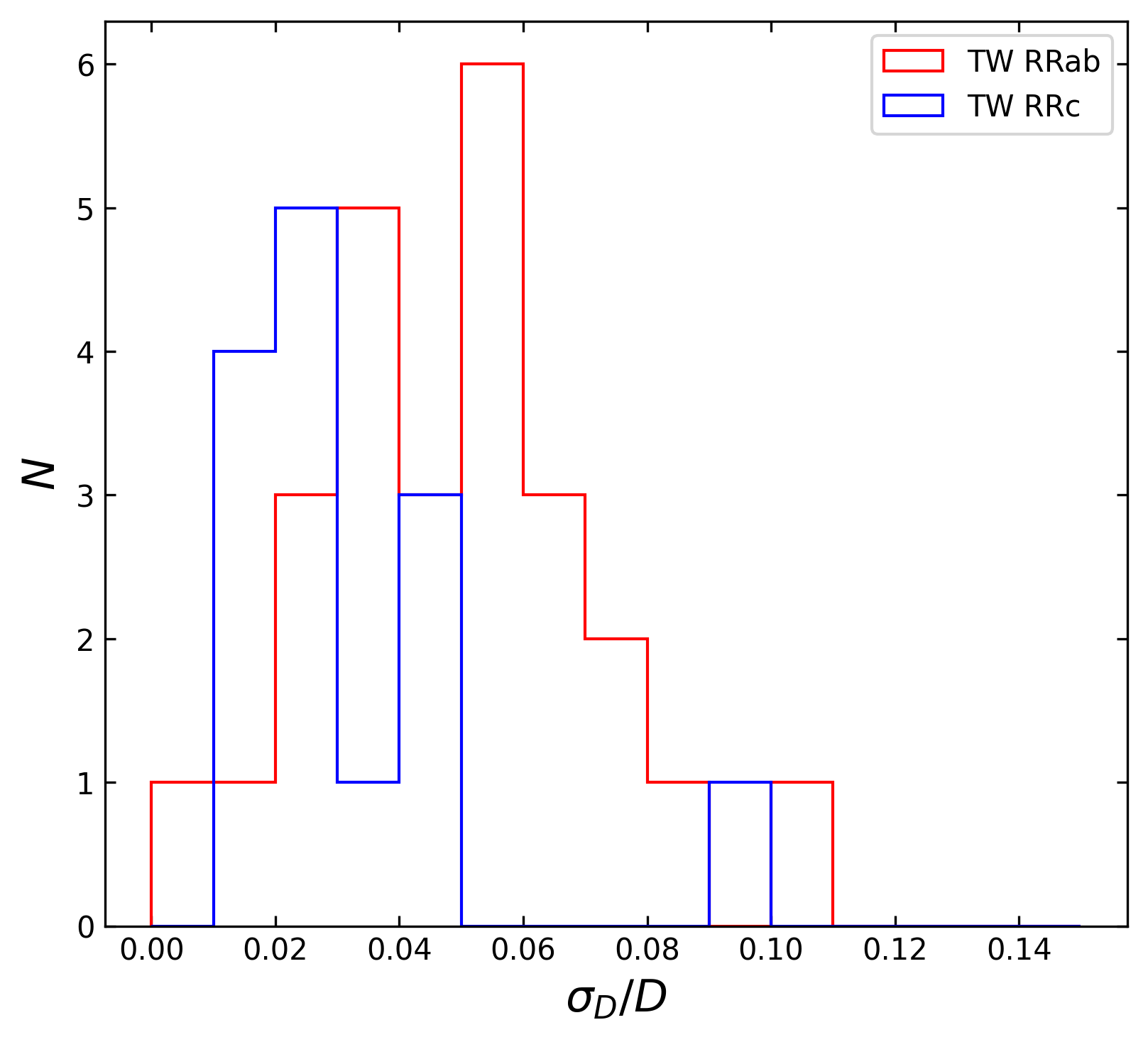}
\caption{{\it Left panel:} Distributions of metallicity uncertainties estimated from members of GCs, with red for type RRab stars, blue for type RRc stars, solid line for this work (TW), and dash-dot for C22. {\it Right panel:} Similar to the left panel but for relative distance uncertainties.}
\end{figure*}

\subsubsection{$M_{G}$--{\rm[Fe/H]} Relations}
We now use the RRL\_CAL\_DIS sample to calibrate the $M_{G}$--[Fe/H] relations for both type RRab and type RRc stars. First, the $G$-band absolute magnitudes of the sample stars are derived from the parallax-based distances and extinction corrections, assuming $R_{G} = 2.516$ \citep{2021ApJ...907...68H}. Uncertainties in  $M_{G}$  are primarily from three sources: the distance errors, the photometric measurement errors of the $G$-band, and an adopted error of $E (B-V)$ = 0.05\,mag. To ensure the high quality of our calibration samples, we further require that $e_{M_{G}} \leq$ 0.15\,mag both for type RRab and type RRc stars. With these cuts, 205 type RRab and 31 type RRc stars remain in our RRL\_CAL\_DIS sample stars. A simple linear relation between $M_{G}$ and [Fe/H] is then adopted:

\begin{equation}
M_{G} = b\,{\rm[Fe/H]}+ c\text{,}
\end{equation}
where $M_{G}$ is the $G$-band absolute magnitude, [Fe/H] is the photometric metallicity estimated from Equations.\,(1) and (2), and $b$ and $c$ are fit coefficients. The resulting fit coefficients for type RRab and type RRc stars are listed in Table\,2.

The fitting results for type RRab and type RRc stars are presented in Figures\,6 and\,7, respectively.
In Figure\,6, we show that the new calibration in this study (black line) is consistent with that of \cite{2018MNRAS.481.1195M}. The corresponding absolute $G$-band magnitudes at [Fe/H]$=-$1.5 are 0.584$\pm 0.036$\,mag and 0.631$\pm 0.103$\,mag for type RRab and type RRc stars, respectively, again in good agreement with the results of \cite{2018MNRAS.481.1195M}. The  scatters of the fitting residuals are only 0.12 and 0.10\,mag for type RRab and type RRc stars, respectively, smaller than the results achieved in most recent calibrations (e.g., \citealt{2018MNRAS.481.1195M}; \citealt{2019MNRAS.490.4254N}).

We apply the newly constructed $M_{G}$--[Fe/H] relations to estimate distances for the type RRab and type RRc stars in our final RRL sample. Generally, the extinction corrections are adopted from the SFD98 map. However, for the LMC and SMC regions, a more precise map, provided by \cite{2021ApJS..252...23S}, is adopted for these corrections. For the low-latitude region with $|b| \leq 25^{\circ}$, the corrections for $E (B-V)$ are those provided by C22, if available, for type RRab stars. For member RRLs of GCs (see details in Section\,4.2), the values of $E (B-V)$ are taken from \citet[][hereafter H10]{2010arXiv1012.3224H}. 

A summary of the extinction corrections is presented in Table\,3. The distances for all our RRL sample stars (115,410 type RRab and 20,463 type RRc stars) are derived in this manner. 
The uncertainties of the distances are contributed by random and method errors.
The random error can be estimated by a MC simulation similar to that mentioned in Section\,3.1.
For each star, 1000 MC simulations are performed by sampling the uncertainties of the fit coefficients, $G$-band magnitude, and photometric metallicity, as well as a fixed uncertainty of 0.05 mag for $E (B-V)$.
The random error of each star is calculated from the distribution yielded from the MC simulations.
The method error is set to 0.12 and 0.10\,mag for type RRab and type RRc stars, respectively, adopted from the dispersion of the calibrated relation mentioned above.   
The typical distance uncertainties are 9.91\% and 9.77\% for type RRab and type RRc stars, respectively.

\begin{figure*}[!htbp]
\includegraphics[width=3.5in]{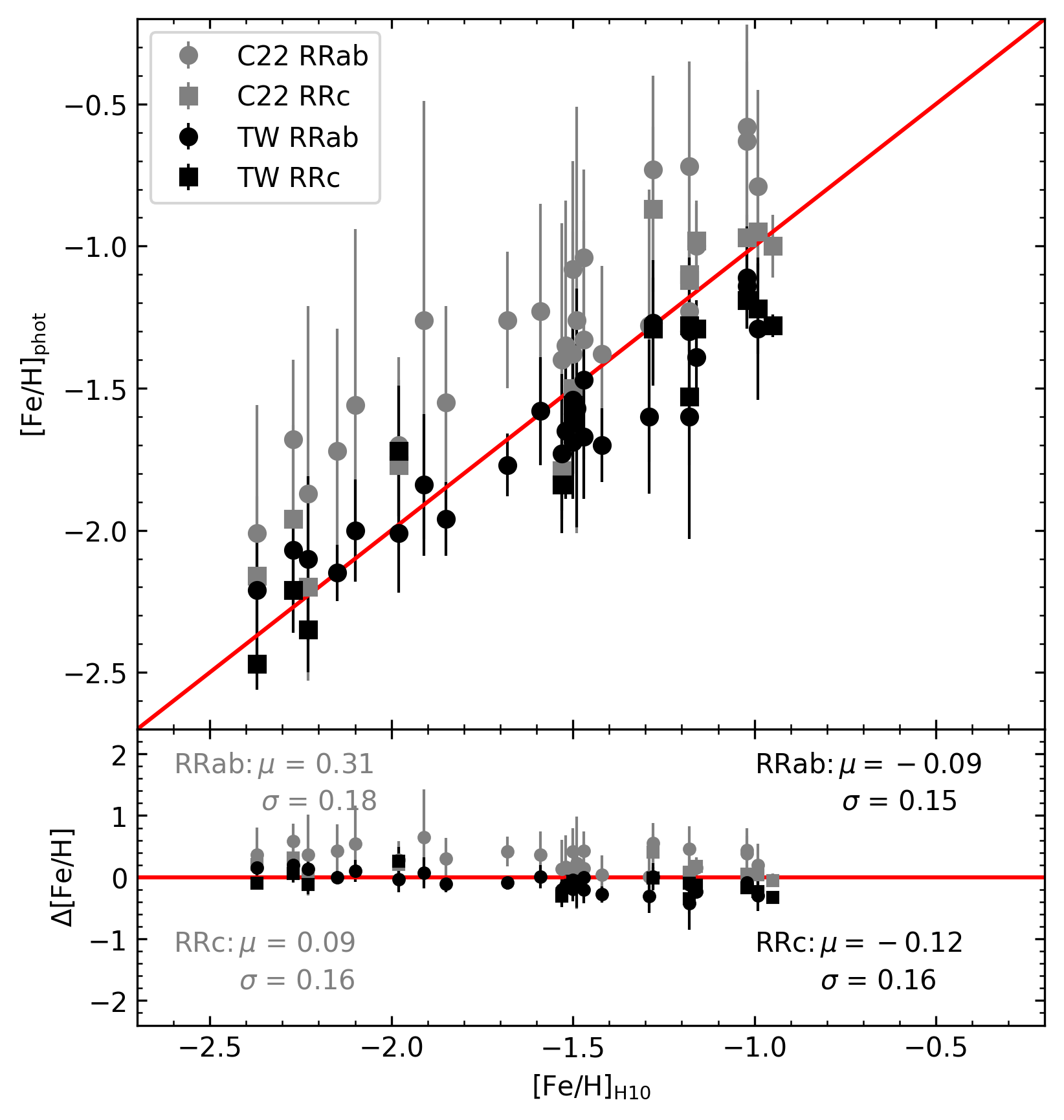}
\includegraphics[width=3.5in]{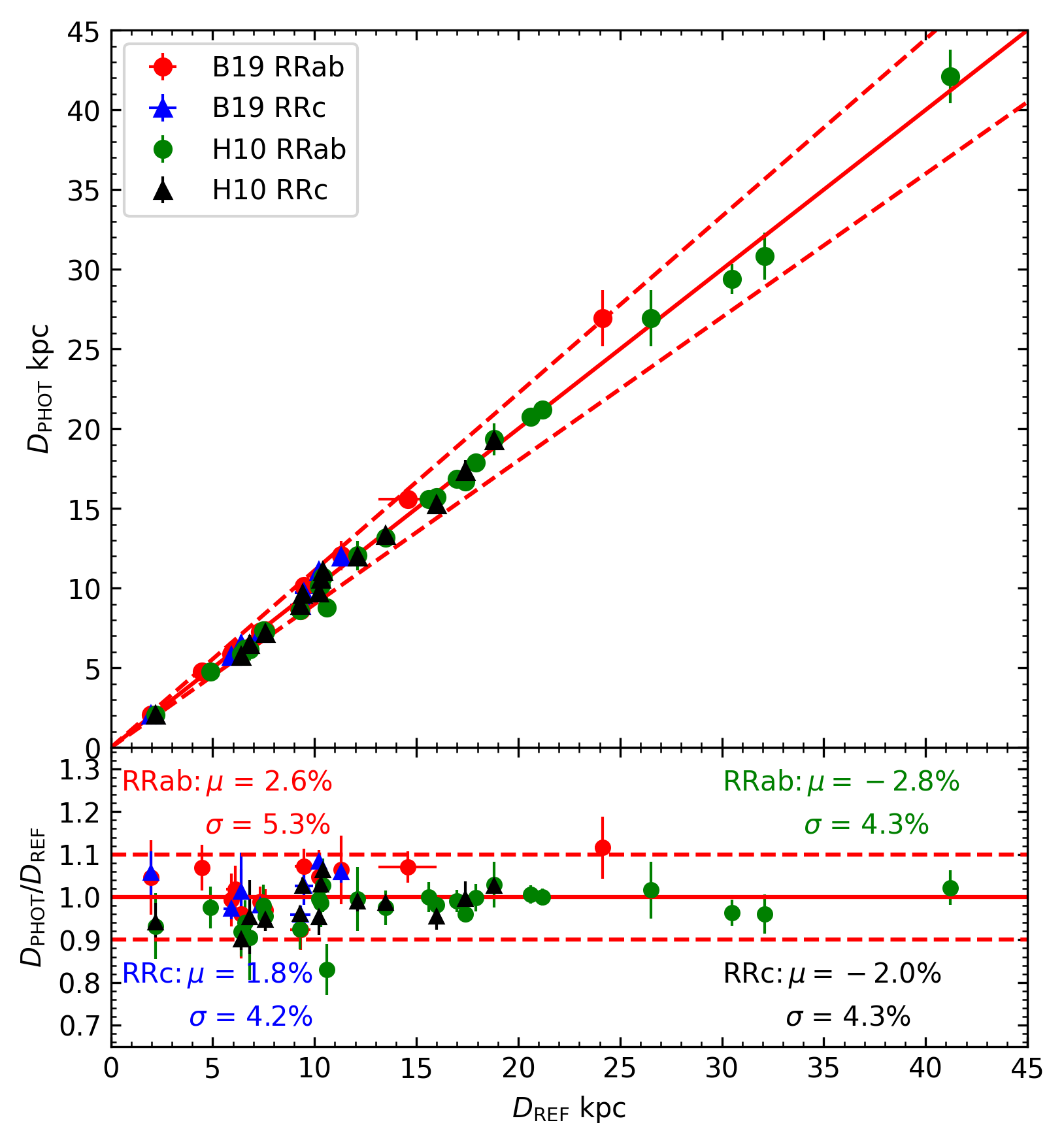}
\caption{{\it Left panel}: Comparison of our estimated photometric metallicities of GCs with those from H10. The black dots and squares indicate the photometric metallicities of GCs derived from type RRab and type RRc stars by our photometric method, respectively. Photometric metallicities of GCs (gray dots and squares) from C22 are also shown for comparison. The metallicity differences {$\rm\Delta[Fe/H]$} (in the sense of this work or C22 minus that of H10) are shown in the lower part of the panel, with the mean and standard deviation of the differences marked. {\it Right panel}: Comparison of estimated distances of GCs with those from B19 and H10. The dots and triangles indicate the distances of GCs derived from type RRab and type RRc stars by our $M_{G}$--[Fe/H] relations, respectively. The red dashed lines mark $D\rm_{PHOT}$ $= 1.1 D\rm_{REF}$ and $D\rm_{PHOT}$ $= 0.9 D\rm_{REF}$, respectively. The ratios $D\rm_{PHOT}$/$D\rm_{REF}$ are shown in the lower part of the panel, with the mean and standard deviation of the relative difference $\Delta D/D$ marked in the four corners.}
\end{figure*}

\subsubsection{Near-infrared $PMZ$ Relations}
The $PMZ$ relations for RRLs have been studied by many authors (e.g., \citealt{2003MNRAS.344.1097B, 2017ApJ...838..107S, 2015ApJ...807..127M,2018MNRAS.481.1195M, 2018MNRAS.480.4138M}) in the near-infrared 2MASS $K\rm_{s}$ and WISE $W1$ passbands. Here, we aim to re-calibrate the $PM_{K{\rm _{s}}}Z$ and $PM_{W1}Z$ relations for type RRab stars using our RRL\_CAL\_DIS sample. The absolute magnitudes at $K_{\rm s}$ and $W1$ are derived for 823 type RRab stars in both bands by assuming $R_{K_{\rm s}} = 0.306$ and $R_{W1} = 0.186$ from \citet{2013MNRAS.430.2188Y}. To ensure the quality of the  $PWZ$ relations, we remove stars with absolute magnitude uncertainties greater than 0.08\,mag in either the $K_{\rm s}$ and $W1$ bands, leaving 159 and 164 type RRab stars, respectively. A simple 2D linear function is adopted for the $PMZ$ relations:

\begin{equation}
M_{K_{\rm s}/W1} = d\,{\rm log}(P) + e\,{\rm[Fe/H]}+ f\text{,}
\end{equation}
where $M_{K_{\rm s}/W1}$ is the $K\rm _{s}$ or $W1$ band absolute magnitude, ${\rm log}(P)$ is the logarithm of the pulsation period, and [Fe/H] is the photometric metallicity, and $d$, $e$, and $f$ are the fit coefficients. The resulting fit coefficients of the $PM_K{\rm_{s}}Z$ and $PM_{W1}Z$ relations are listed in Table\,2; the fits are shown in Figures\,8 and\,9, respectively. Our PLZ relations yield a residual RMS of 0.14 and 0.09 mag in the $K_{\rm S}$ and $W_1$ bands, again better than the results of most recent efforts (e.g., \citealt{2018MNRAS.481.1195M}; \citealt{2019MNRAS.490.4254N}).
 For the $PM_K{\rm_{s}}Z$ relation, the slope of the pulsation period is consistent with those from previous empirical calibration studies \citep{2006MNRAS.372.1675S,2008MNRAS.384.1583S,2015ApJ...807..127M,2018MNRAS.481.1195M,2019MNRAS.490.4254N,2021MNRAS.502.4074M} and those given by theoretical studies \citep{2004ApJS..154..633C}. The relation exhibits a mild metallicity dependence, which is again in agreement with the results of previous empirical calibrations \citep{2018MNRAS.481.1195M,2019MNRAS.490.4254N} theoretical studies \citep{2004ApJS..154..633C,2015ApJ...808...50M}. 
For the $PW1Z$ relation, the slope of the period term is consistent with previous results \citep{2014MNRAS.439.3765D,2017ApJ...838..107S,2018MNRAS.481.1195M,2019MNRAS.490.4254N,2021MNRAS.502.4074M}, and the coefficient of the metallicity term agrees with the results of  \cite{2018MNRAS.481.1195M} and \cite{2019MNRAS.490.4254N}. The corresponding $K\rm _{s}$- and $W1$-band absolute magnitudes at [Fe/H] $=-$1.5 and pulsation period $P=$ 0.5238\,day are presented in Table\,2. These values are consistent with those given by \cite{2018MNRAS.481.1195M} and \cite{2021MNRAS.502.4074M}.

\section{Validation of Metallicity and Distance Estimates}

In this section, the photometric-metallicity estimates in this study are compared to other photometric and spectroscopic estimates from the literature. 

\subsection{Validation of Photometric Metallicity from Other Spectroscopic Samples}

Recently, \citet[][hereafter D22]{2022ApJS..261...33D} have derived photometric metallicities for nearly 60,000 type RRab stars directly from their light curves provided by Gaia DR2 using a deep learning technique.
The metallicity of the training sample is adopted from the $I$-band photometric estimates, which are well-calibrated by stars with metallicities measured from high-resolution spectroscopy (HRS) \citep{2021ApJ...920...33D}.
Here, our photometric metallicities of type RRab stars are compared to those from D22 based on 49,969 stars in common, with the result shown in  Figure\,10(a). 
The comparison shows a mild offset of 0.17\,dex (this work minus D22), and a scatter of 0.19\,dex, implying reasonable consistency between this work and D22.
The offset is mainly from the bias of D22's metallicity scale, if we trust the scale of metallicity measured from high-resolution spectroscopy (HRS; see Fig.\,11).

By using 84 stars with metallicity determined by \citet[][hereafter L94]{1994AJ....108.1016L} from low-to-moderate resolution spectra, \citet[][hereafter IB21]{2021MNRAS.502.5686I} re-calibrated the linear $P$--$\phi_{31}$--[Fe/H] relation for type RRab stars with the period and $\phi_{31}$ from Gaia\,DR2 \citep{2019A&A...622A..60C}.
They also re-calibrated the same relation for type RRc stars  by using GC members with metallicity taken from H10.
We then compare metallicities of this work to those yielded by the relations of IB21 for 115,410 type RRab stars (middle panel of Fig.\,10) and 20,463 type RRc stars (right panel of Fig.\,10 released in Gaia\,DR3. 
In general, our photometric estimates agree very well with those of IB21, with an overall scatter around 0.16\,dex.
However, we note that the metallicity scale of IB21 is systematically higher that this work, by about 0.1\,dex.
Moreover, a significant systematic trend along [Fe/H] is detected for both type RRab and type RRc stars.
The offset (IB21 minus this work) is about $-0.5$\,dex at [Fe/H] around $0.5$ and about $0.5$\,dex at [Fe/H] around $-3.0$ for type RRab stars.
For type RRc stars, the offset is minor at metal-rich range while significant up to $0.4$\,dex at the metal-poor end ([Fe/H]$\sim -3.0$).
From checks with HRS measurements (see Fig.\,11), the overall offset and the systematic trend, as found by our comparisons, are mainly due to the calibrations of IB21.

\citet[][hereafter D21]{2021ApJ...920...33D} presented 
a bibliographical compilation of 183 RRab and 49 RRc stars with metallicity measurements from high-resolution spectroscopy, covering the range $-3.1 \lesssim$ [Fe/H] $\lesssim$0.2. 
All the metalliities are carefully calibrated to the reference scale established by \citet{2021ApJ...908...20C}.
These stars are therefore used to examine the performances of our photometric estimates of metallicity, as well as those of D22 and IB21.
The comparisons are presented in Fig.\,11.
For type RRab stars, the metallicities from this work are in excellent agreement with those of HRS, with a negligible offset of $-0.01$\,dex and a dispersion of 0.24\,dex.
The metallicities of type RRc stars from this work again agree very well with those of HRS, with a scatter of only 0.16\,dex, although there exists a mild offset of $0.16$\,dex (this work minus HRS).
The D22 results for type RRab stars are also consistent with those from HRS, with a small dispersion of 0.21\,dex, while the scale of D22 is slightly lower than that of HRS, by $0.11$\,dex. 
The metallicity comparisons between IB21 and HRS exhibit large scatters of 0.27/0.33\,dex and offsets of 0.09/0.19\,dex (IB21 minus HRS) for type RRab/c stars.
Moreover, the metallicity differences show a significant trend with [Fe/H] (see the right two panels of Fig.\,11).

\begin{figure}[!htbp]
\begin{minipage}[t]{0.5\linewidth}
\centering
\includegraphics[width=3.35in]{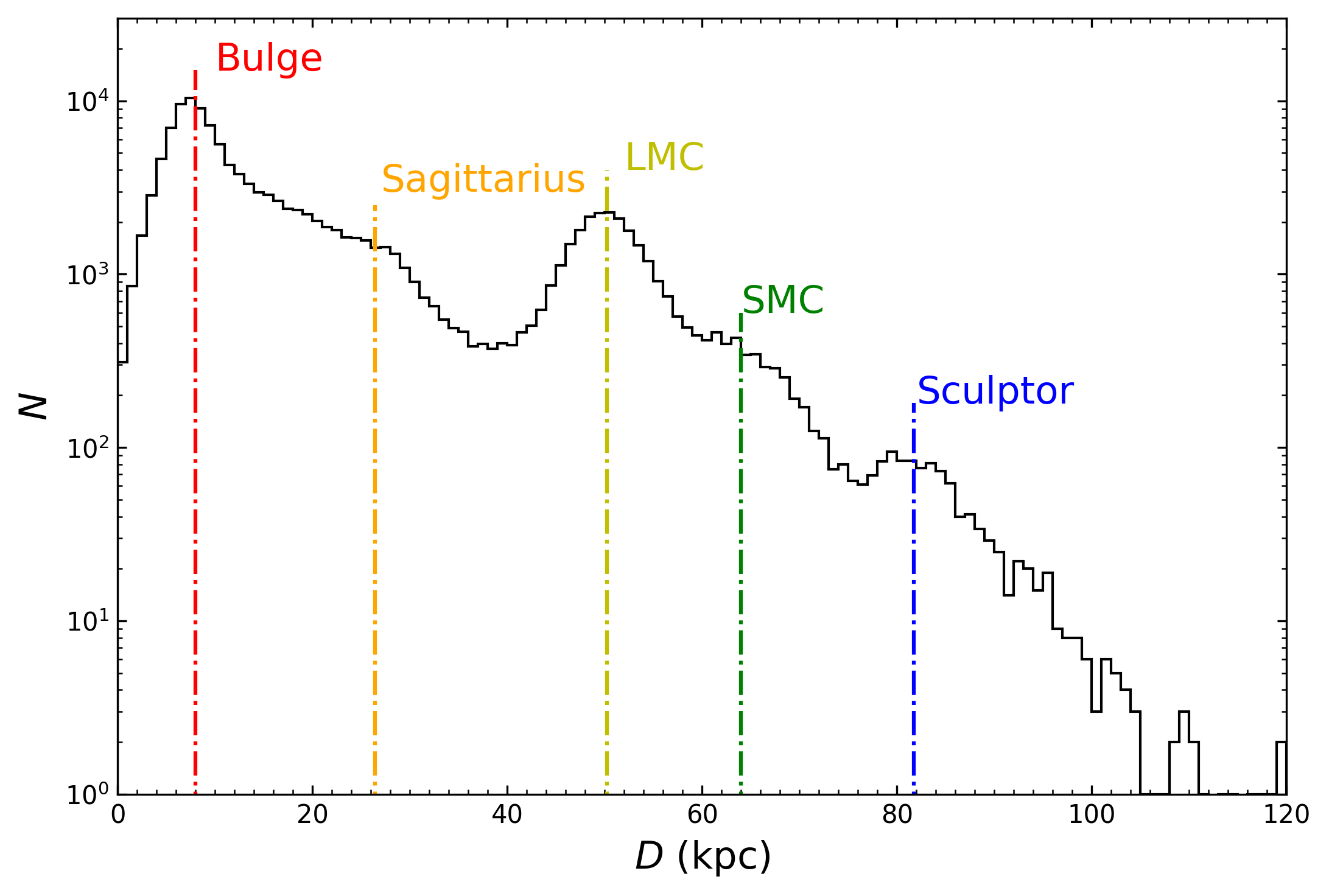}
\end{minipage}
\caption{Histogram of the distances for 115,410 type RRab and 20,463 type RRc stars. The red, orange, yellow, green, and blue dashed lines mark the positions of the Galactic bulge, Sagittarius dwarf galaxy, LMC, SMC, and Sculptor dwarf galaxy, respectively.}
\end{figure}

\begin{figure*}[!htbp]
\begin{minipage}[t]{0.5\linewidth}
\centering
\includegraphics[width=7.in]{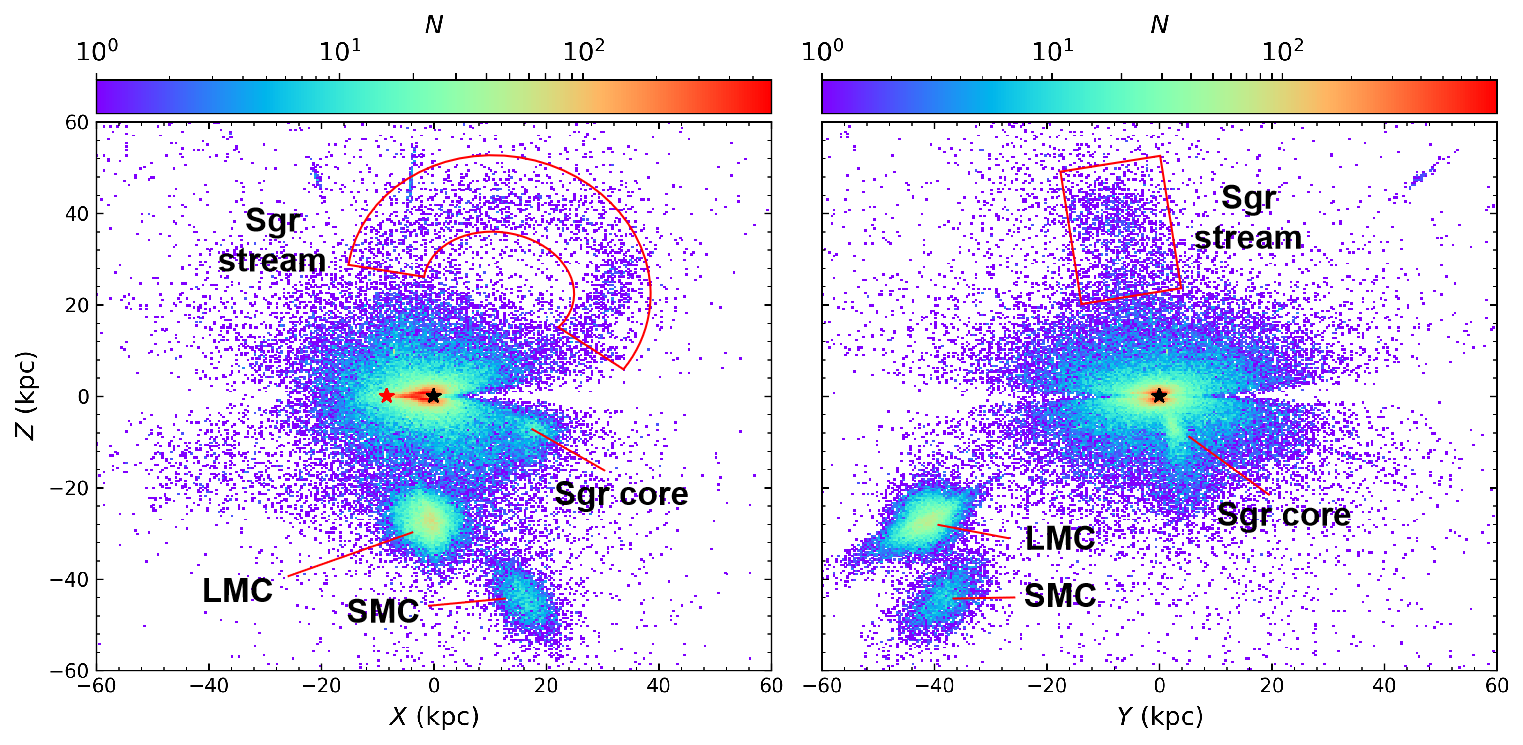}
\end{minipage}
\caption{Spatial distribution of our final RRL sample stars in the $X$--$Z$ (left) and $Y$--$Z$ (right) planes. The Galactic center and Sun are represented by the black and red stars, respectively. The stellar number densities (in a bin size of 0.4\,kpc for both axes) are represented by the color bars shown.
The well known satellites and structures, e.g. LMC, SMC, and Sgr core and stream, are clearly seen.}
\end{figure*}

\subsection{Validation with Globular Clusters}
The member stars of globular clusters (GCs) are expected to be born at the same time and with almost the same metallicities. The RRLs of GCs are therefore selected in this study for testing the accuracies of our photometric metallicities and distance estimates from our newly derived $M_{G}$--[Fe/H] relations. Prior to these comparisons, we apply the following cuts to select member stars of GCs from our RRL sample:
\begin{itemize}[leftmargin=*]

\item The positions must within 15 half-light radii $r_{\rm h}$ ( from H10) from the center of the GC

\item The proper motions must satisfy $|\mu_{\alpha} - \mu_{\alpha, {\rm GC}}| \leq 8\, \sigma_{\mu_{\alpha, {\rm GC}}}$ and\,$|\mu_{\delta} - \mu_{\delta, {\rm GC}}| \leq 8\, \sigma_{\mu_{\delta, {\rm GC}}}$

\end{itemize}
The above cuts are applied to 157 GCs in H10. The proper motions and uncertainties for each GC are taken from \citet{2021MNRAS.505.5978V}.  

Finally, to ensure our estimates of the mean metallicity and distance of GCs  are reasonably well-determined from  by our RRL sample stars, a minimum number of 5 type RRab and 3 type RRc stars are required for selecting members of each GC. In this way, 431 type RRab and 78 type RRc stars are selected from 26 and 14 GCs, respectively. 

The mean metallicity (both for this study and C22) and distance, as well as their uncertainties (given by the standard deviations of the distributions), are estimated from the distributions of selected member RRLs for each GC (e.g., see Figure\,12). 
The results are listed in Table\,4.
As shown in Figure\,13, the typical metallicity uncertainties in this study are 0.21\,dex and 0.11\,dex for type RRab and type RRc stars, respectively, significant smaller than the values of 0.38\,dex and 0.13\,dex found for C22.

The relative distance errors are mostly within 8\% and 5\% for type RRab and type RRc stars, with medians of 4.9\% and 2.4\%, respectively.
The small uncertainties revealed by the GCs, even smaller than the typical ones of the full sample, are mainly due to the nearly nil extinction corrections.

Figure\,14 shows a comparison between the metallicities of GCs from our photometric methods and those from H10. 
The results are in excellent agreement with those of H10, with a negligible offset of $-$0.09\,dex and a small scatter of 0.15\,dex for type RRab stars, and a mild offset of $-0.12$\,dex and a small scatter of 0.16\,dex for type RRc stars. 
In contrast, the difference between photometric metallicities for type RRab stars from C22 and those from H10 exhibit a significant offset of $0.31$\,dex and a relatively large scatter of 0.18\,dex. 
The photometric metallicities from C22 for type RRc stars are slightly higher than that of the GCs, by 0.09\,dex, with a moderate scatter of 0.16\,dex.
In summary, the accuracies of the photometric metallicities obtained in this study are much better than those achieved by C22 (especially for type RRab stars).

For distances, we compare our results to those of \citet[][hereafter B19]{2019MNRAS.482.5138B}, which determine accurate kinematic distances for 53 GCs by fitting the proper motion and line-of-sight velocity dispersion profiles to N-body simulations for individual GCs. The comparison is shown in Figure 14. From inspection, our derived distances for GCs are in excellent agreement with the kinematic distances of B19, with a negligible median offset of the relative distance difference $\Delta D/D$ of 2.6\%/1.8\%, and a small scatter of 5.3\%/4.2\% for type RRab/RRc stars, respectively.
In addition, our distances are also consistent with the previous work of H10, exhibiting a negligible offset of the relative distance difference $\Delta D/D$ of $-$2.8\%/$-$2.0\%, and a small scatter of 4.3\%/4.3\% for type RRab/RRc stars, respectively. We also note that, for NGC 1851, IC 4499, NGC 6121, NGC 6171, and NGC 6362, the distances derived from RRc stars are in excellent agreement with the results of type RRab stars, with the relative distance differences smaller than a few per cent. 

\begin{figure}[!htbp]
\begin{minipage}[t]{0.5\linewidth}
\centering
\includegraphics[width=3.4in]{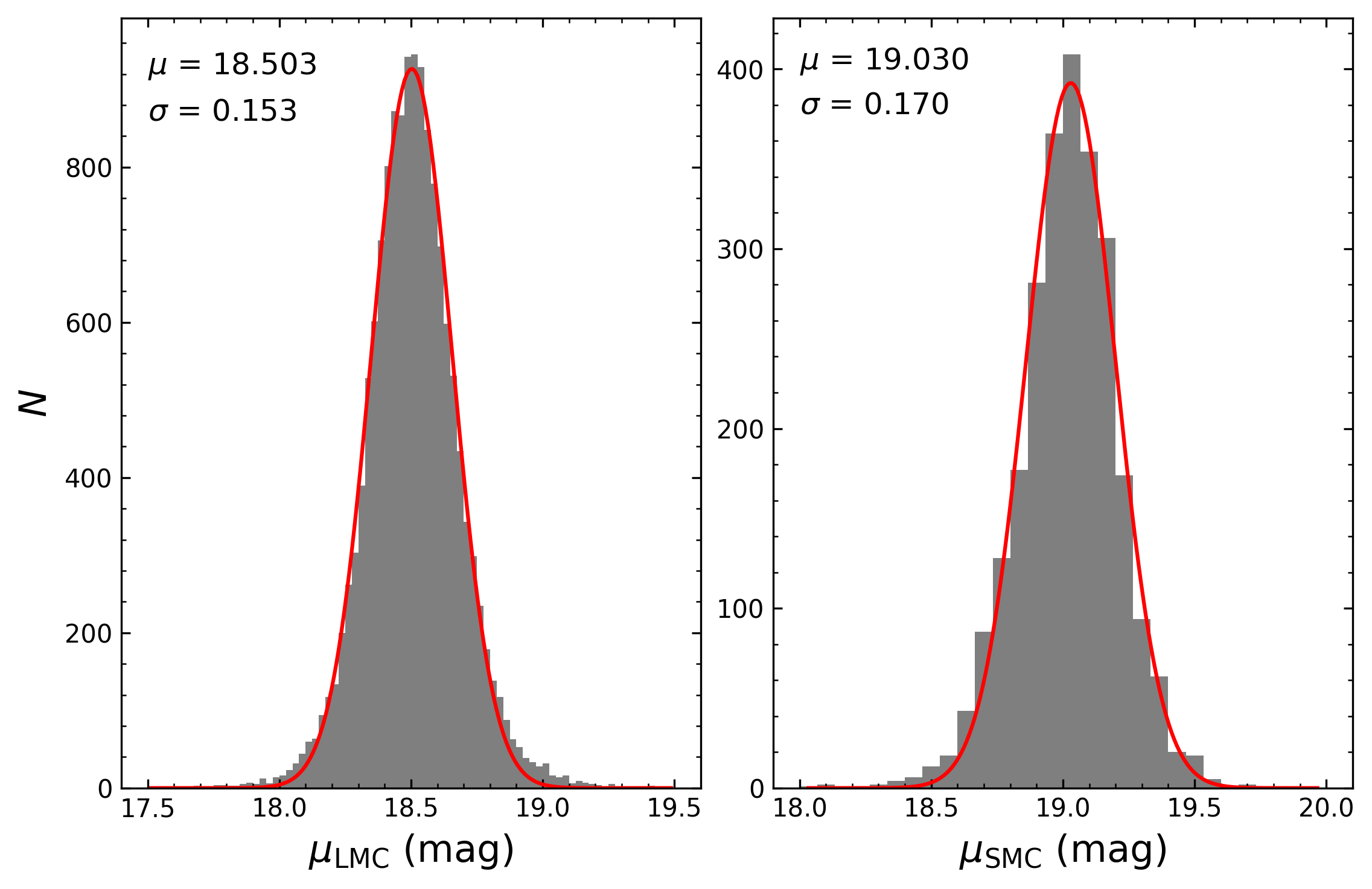}
\end{minipage}
\caption{{\it Left panel}: Histogram of the distance moduli for all type RRab stars selected in LMC. The red line is a Gaussian fit to the distribution, with the mean and dispersion of the Gaussian marked in the top-left corner. {\it Right panel}: Same as the left panel, but for the SMC.}
\end{figure}
\section{The RR Lyrae Sample}

In this section, we describe the properties and potential applications of our final RRL star sample. 

Our final RRL sample contains 135,873 RRLs (115,410 type RRab and 20,463 type RRc stars), with precise metallicity and distance estimates from the newly constructed $P$--$\phi_{31}$--$R_{21}$--[Fe/H], $P$--$R_{21}$--[Fe/H] and $M_{G}$--[Fe/H] relations. The distance distribution of the final RRL sample is shown in Figure\,15, with five prominent peaks at heliocentric distances of 8, 26, 50, 64, and 82\,kpc, corresponding to the positions of the Galactic bulge, Sagittarius dwarf galaxy, LMC,  SMC, and Sculptor dwarf galaxy, respectively. The sample stars can reach as far as 100\,kpc, but most of them are within 30\,kpc, except those belonging to the LMC/SMC. We then calculated the 3D positions of our sample stars from their sky positions ($l, b$) and distances. We use a right-handed Cartesian Galactocentric coordinate system ($X$, $Y$, $Z$), with $X$ pointing towards the Galactic center, $Y$ in the direction of Galactic rotation, and $Z$ towards the north Galactic pole. The position of the Sun is set to ($X$, $Y$, $Z$) = ($-$8.34, 0.00, 0.00)\,kpc \citep{2014ApJ...783..130R}. 

The spatial distributions of our sample stars in the $X$--$Z$ and $Y$--$Z$ planes are presented in Figure\,16. The sample covers a large halo volume of $|X|\leq30$\,kpc, $|Y|\leq30$\,kpc and $|Z|\leq30$\,kpc,  with a sufficient number of stars to explore questions related to Galactic structure, and the nature of the stellar populations in this region. Table\,5 lists the columns included in the final online sample catalog, which is also available at \url{http://doi.org/10.5281/zenodo.6731860}.

This sample will clearly be very useful to study the structure, chemical, and kinematic properties of the Galactic halo. For example, the sample has already been applied to find substructures and study their chemical properties (Wang et al. submitted). To show the potential power of this sample, we use it derive the distance moduli and consider the chemical properties of LMC and SMC, as an example. 

\begin{figure*}[!htbp]
\begin{minipage}[t]{0.5\linewidth}
\centering
\includegraphics[width=7.in]{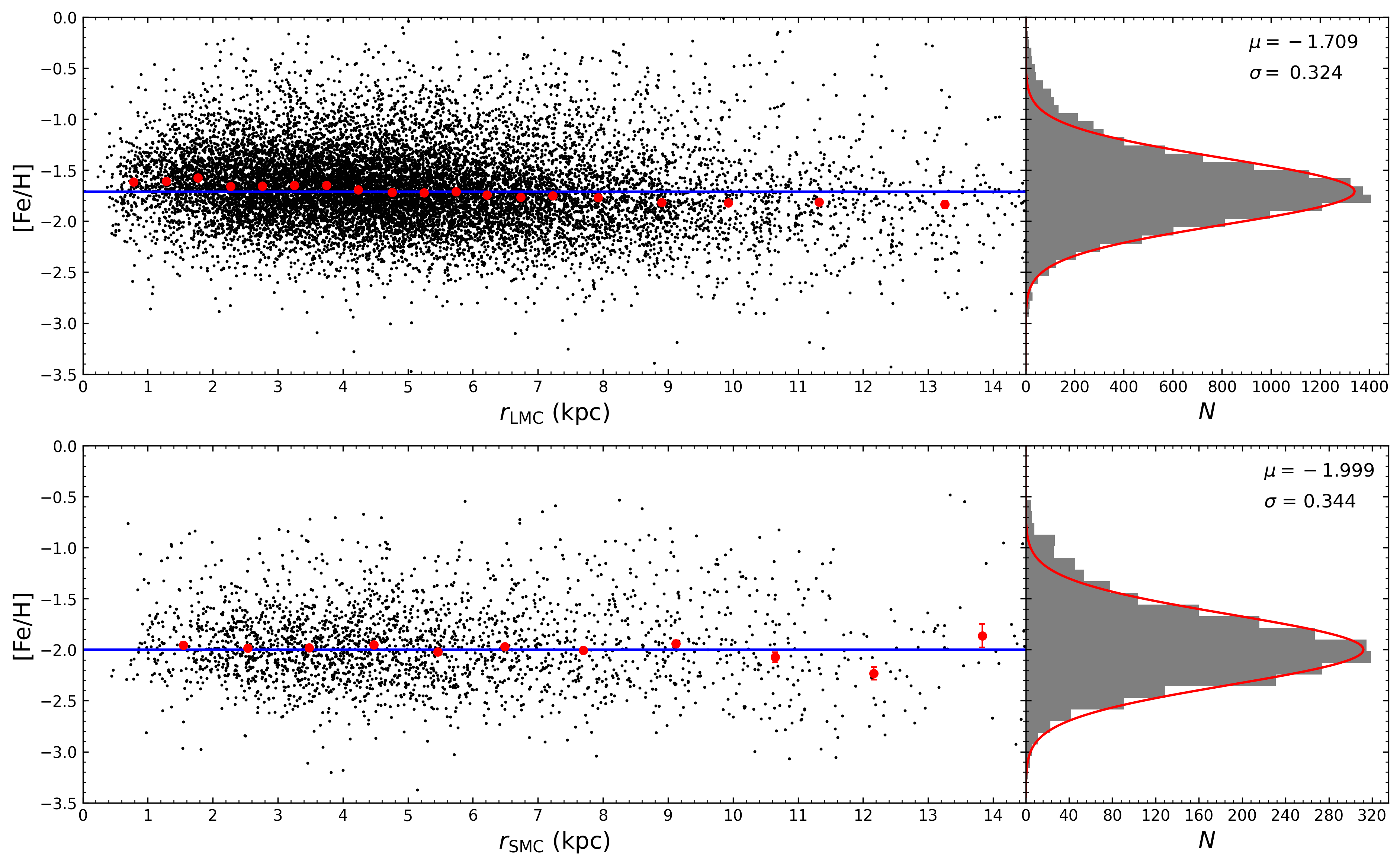}
\end{minipage}
\caption{{\it Top panel}: Distribution of metallictiy, as a function of the radial distance from the LMC center ({\it left}), and a histogram of the metallicity for 14,242 type RRab stars in the LMC ({\it right}). The mean metallicity and the values of the standard deviation derived by a Gaussian fit (red line) are shown in the top-right part of right panel. The mean metallicity is shown on the left by the blue line. Large red circles mark the median metallicity values in each radial bin, with bin size in $r_{\rm LMC}$ of 0.5\,kpc for $1\leq r_{\rm LMC}\leq 7.5$\,kpc, 1\,kpc for $r_{\rm LMC}< 1$\,kpc and $7.5 < r_{\rm LMC} < 10.5$\,kpc, and 2\,kpc for $r_{\rm LMC}\geq 10.5$\,kpc. {\it Bottom panel}: Same as top panel, but for 2446 type RRab stars, with the bin size in $r_{\rm SMC}$ of 1\,kpc for  $2\leq r_{\rm SMC}\leq 7$\,kpc, 1.5\,kpc for $r_{\rm SMC}> 7$\,kpc, and 2\,kpc for $r_{\rm SMC}< 2$\,kpc.}
\end{figure*}

The LMC and SMC are the two largest satellite galaxies of the Milky Way, located at distances of about 50 and 60\,kpc from the Sun, respectively. They are regarded as the cornerstones of the cosmological distance ladder, and it is therefore of vital importance to obtain very accurate and consistent distances for the two systems by different tracers using of different methods. Here, we derive the distances for the LMC and SMC from the type RRab stars in our final sample. Member stars of the LMC and SMC are selected by requiring that our type RRab stars lie within a radius of $20^{\circ}$ and $10^{\circ}$ from the centers of the two systems, respectively. The positions of the centers of the LMC and SMC are $(\alpha_{\rm C,LMC}, \delta_{\rm C,LMC})=(81^\circ.28, -69^\circ.78)$, defined by \cite{2001AJ....122.1827V} and $(\alpha_{\rm C,SMC}, and \delta_{\rm C,SMC})=(12^\circ.80, -73^\circ.15)$,  defined by \cite{2000A&A...358L...9C}, respectively. In this way, 15,038 and 2696 type RRab stars with distance errors smaller than 12\% are selected in the LMC and SMC, respectively. Distance moduli of $\mu_0 = 18.503 \pm 0.001(stat) \pm 0.040(syst)$\,mag and $\mu_0 = 19.030 \pm 0.003(stat) \pm 0.043(syst)$ are derived for the LMC and SMC, respectively, from the selected type RRab stars (see Figure\,17). 
A summary of our error budget is available in Table\,6. 
The standard deviations of the distance distributions are 0.153 and 0.170\,mag, respectively, for the LMC and SMC, which are mainly contributed by the physical extensions of the two systems themselves.
The results are in excellent agreement with previous measurements from tracers such as Cepheids (e.g., \citealt{2012ApJ...758...24F,2016ApJ...816...49S}), RRLs \citep{2019MNRAS.490.4254N}, eclipsing binaries \citep{2019Natur.567..200P,2020ApJ...904...13G}, and the reported mean values of different tracers \citep{2014AJ....147..122D,2015AJ....149..179D}.

We also consider the question whether there exist metallicity gradients in the LMC or SMC. We employ the following transformation equations \citep{2001ApJ...548..712W,2001AJ....122.1807V} to calculate the three-dimensional coordinates for each type RRab star in the LMC or SMC,
\begin{equation}
\begin{split}
x =\,& -d\times {\rm cos}(\delta){\rm sin}\, (\alpha-{\rm \alpha _{C}})\text{,}\\
y =\,& d\times [{\rm sin(\delta)\, cos(\delta _{C})} \\
&- \rm{ cos(\delta)\, sin(\delta _{C})\, cos(\alpha - \alpha _{C})}]\text{,}\\
z =\,& d\times {\rm[cos(\delta)\, cos(\delta _{C})\, cos(\alpha - \alpha _{C})}\\
&  + {\rm sin(\delta)\, sin(\delta _{C})]-D_{C}}\text{.}
\end{split}
\end{equation}
where $d$ is the heliocentric distance of each type RRab star, $\alpha\rm _{C}$ and $\delta\rm _{C}$ are the equatorial coordinates of the LMC/SMC centers, and $D\rm _{C}$ is the mean distance of the LMC/SMC, given by our determinations above, i.e., $D_{\rm C,LMC} = 50.19$\,kpc for the LMC and $D_{\rm C,SMC} = 63.97$\,kpc for the SMC. Then, the radial distances from the LMC/SMC center, $r=\sqrt{x^2+y^2+z^2}$, are derived for the type RRab stars in the LMC/SMC. To obtain a reliable metallicity trend with radial distance $r$, we apply the following cuts to the selected type RRab stars: i) the photometric metallicity lies in the range of $-3.5 \le$\,[Fe/H]\,$\leq 0.5$ and with uncertainties smaller than 0.5\,dex, ii) the distance errors smaller than 12\% , and iii) $r \leq 14.5$\,kpc. After applying these cuts, 14,242 and 2446 type RRab stars are left in LMC and SMC, respectively. 

The metallicity distributions, as a function of $r$, for the LMC/SMC are shown in Figure\,18. For the LMC, the mean value of metallicity is [Fe/H]$=-1.709 \pm 0.003(stat) \pm 0.023(syst)$, in agreement with the estimate of [Fe/H]\,=\,$-1.59 \pm 0.31$ by \citet[][hereafter S16]{2016AcA....66..269S}. Generally, the metallicity distribution shows a negative slope for $r \leq 5$\,kpc, then tends to be flat for $5 < r < 7$\,kpc, and then again presents a negative gradient for $r > 7$\,kpc. Within the inner 5\,kpc, we find a metallicity gradient of $-0.029 \pm 0.004$\,dex\,kpc$^{-1}$, steeper than found by previous studies using RR Lyrae stars \citep{2010MNRAS.408L..76F,2013MNRAS.431.1565W}, but consistent with S16's result within 1$\sigma$. The metallicity gradient is nearly flat around $r_{\rm LMC}=6$\,kpc. The slope of the gradient is $-0.012 \pm 0.005$\,dex\,kpc$^{-1}$ for $r$ beyond 7\,kpc. Excluding  the last bin with large uncertainties, the slope of the gradient becomes $-0.018 \pm 0.007$\,dex\,kpc$^{-1}$, smaller than the result of $-0.047 \pm 0.003$\,dex\,kpc$^{-1}$ from \cite{2009A&A...506.1137C} using asymptotic giant branch (AGB) stars in the same region. The overall slope is $-0.024 \pm 0.001$\,dex\,kpc$^{-1}$ for the entire LMC, consistent with S16's value of $-0.019 \pm 0.002$\,dex\,kpc$^{-1}$. For the SMC, the mean value of metallicity is [Fe/H]$=-1.999\pm 0.008(stat) \pm 0.023(syst)$, also in agreement with S16's result (${\rm [Fe/H]}=-1.85 \pm 0.33$). We find that the SMC exhibits a very weak metallicity gradient of $-0.007 \pm 0.003$\,dex\,kpc$^{-1}$. The trend is similar to previous results derived from the AGB stars, stellar clusters, and type RRab stars \citep{2009A&A...506.1137C,2009AJ....138..517P,2015MNRAS.449.2768D,2016AcA....66..269S}.

\section{Summary}

In this paper, using a training sample of 2687 RRLs (2046 type RRab and 641 type RRc stars) with accurate pulsation periods, $P$, Fourier parameters, $\phi_{31}$, $R_{21}$, and metallicity, [Fe/H], measurements from the $Gaia$, LAMOST, and SDSS surveys, new $P$--$\phi{31}$--$R_{21}$--[Fe/H] and $P$--$R_{21}$--[Fe/H] relations are derived for type RRab and type RRc stars, respectively. Precise photometric metalliciies for the 135,873 RRLs (115,410 type RRab and 20,463 type RRc stars) are derived from the constructed relations. 
Comprehensive external checks demonstrate no significant offsets between our photometric metallicities; the typical precisions are 0.24 and 0.16\,dex for type RRab and type RRc stars, respectively. From about one thousand local bright RRLs with photometric meatallicity estimates and accurate distance estimates from $Gaia$ parallaxes, the $G$-band absolute magnitude-metallicity relations and near-infrared period-absolute magnitude-metallicity ($PLZ$) relations are re-calibrated. 
With the newly constructed $M_{G}$--[Fe/H] relations, the distances for all of the 135,873 RRLs with photometric-metallicity estimates are derived, with a typical uncertainty of 9-10\% (depending on the errors of extinction corrections).

To show the power of this sample, we present distance measurements and chemical studies of the LMC and SMC as one example. Using over ten thousand  type RRab stars, distance moduli of $\mu_0 = 18.503 \pm 0.001(stat) \pm 0.040(syst)$\,mag and $\mu_0 = 19.030 \pm 0.003(stat) \pm 0.043(syst)$\,mag are found for the LMC and SMC, respectively, consistent with previous measurements. In addition, a mean metallicity of [Fe/H]$=-1.709 \pm 0.003(stat) \pm 0.023(syst)$ for the LMC and [Fe/H]$=-1.999 \pm 0.008(stat) \pm 0.023(syst)$ for the SMC are found, again in agreement with the results from previous studies. Moreover, we identify a mild gradient of $-0.024 \pm 0.001$\,dex\,kpc$^{-1}$ for the entire LMC, while the SMC exhibits a nearly flat metallicity gradient of $-0.007 \pm 0.003$\,dex\,kpc$^{-1}$.

\section*{Acknowledgements} 
We thank an anonymous referee for helpful comments.
This work is supported by National Natural Science Foundation of China grants 11903027, 11973001, 11833006, U1731108, 12090040, 12090044 and National Key R \& D Program of China No. 2019YFA0405500. Y.H. and X.Y.L. are supported by the Yunnan University grant No. C176220100006 and 2021Z015, respectively. T.C.B. acknowledges partial support for this work from grant PHY 14-30152; Physics Frontier Center/JINA Center for the Evolution of the Elements (JINA-CEE), awarded by the US National Science Foundation. H.W.Z. acknowledges the science research grants from the China Manned Space Project with No. CMS-CSST-2021-B03.

This work has made use of data from the European Space Agency (ESA) mission $Gaia$ (https://www.cosmos.esa.int/gaia), processed by the $Gaia$ Data Processing and Analysis Consortium (DPAC, https://www.cosmos.esa.int/web/gaia/dpac/consortium). 

The Guoshoujing Telescope (the Large Sky Area Multi-Object Fiber Spectroscopic Telescope, LAMOST) is a National Major Scientific Project built by the Chinese Academy of Sciences. Funding for the project has been provided by the National Development and Reform Commission. LAMOST is operated and managed by the National Astronomical Observatories,
Chinese Academy of Sciences.

\bibliographystyle{apj}
\bibliography{RRLGaiaDR3}

\end{document}